\documentclass[preprint,12pt, authoryear]{elsarticle}

\usepackage[english]{babel}
\usepackage{multirow}
\usepackage{amsmath}
\usepackage{graphicx}
\usepackage{hyperref}  
\usepackage{enumitem}
\usepackage{a4wide}
\usepackage{booktabs}
\usepackage{amssymb}
\usepackage{comment}
\usepackage{subcaption}
\usepackage{placeins} 
\usepackage[super]{nth}

\usepackage{comment}
\usepackage{eurosym}
\usepackage{tikz}
\usepackage{collcell}
\usepackage{url}
\usepackage[toc,page]{appendix}

\usepackage[utf8]{inputenc}
\usepackage[normalem]{ulem}
\usepackage{xcolor}

\newcommand{\aw}[1]{{\color{black}#1}}
\newcommand{\er}[1]{{\color{blue}#1}}
\newcommand{\fs}[1]{{\color{black}#1}}

\newcommand{\awout}[1]{}
\newcommand{\erout}[1]{}
\newcommand{\fsout}[1]{}

\renewcommand{\aw}[1]{#1}
\renewcommand{\er}[1]{#1}
\renewcommand{\fs}[1]{#1}

\newcommand{\revTwo}[1]{{\color{blue}#1}}
\newcommand{\revTwoOut}[1]{{\color{red}\sout{#1}}}
\renewcommand{\revTwo}[1]{#1}
\renewcommand{\revTwoOut}[1]{}

\newcommand*{\MinNumber}{3.0}%
\newcommand*{\MidNumber}{5} %
\newcommand*{\MaxNumber}{11.0}%
\newcommand*{\Ten}{10.0}%

\newtoggle{inTableHeader}
\toggletrue{inTableHeader}
\newcommand*{\StartTableHeader}{\global\toggletrue{inTableHeader}}%
\newcommand*{\EndTableHeader}{\global\togglefalse{inTableHeader}}%

\let\OldTabular\tabular%
\let\OldEndTabular\endtabular%
\renewenvironment{tabular}{\StartTableHeader\OldTabular}{\OldEndTabular\StartTableHeader}%

\newcommand{\ApplyGradient}[1]{%
  \iftoggle{inTableHeader}{#1}{
    \ifdim #1 pt > \MidNumber pt
        \pgfmathsetmacro{\PercentColor}{max(min(100.0*(#1 - \MidNumber)/(\MaxNumber-\MidNumber),100.0),0.00)} %
        \ifdim #1 pt > \Ten pt
            \hspace{-0.33em}\colorbox{red!\PercentColor!yellow}{#1}
        \else
            \hspace{-0.33em}\colorbox{red!\PercentColor!yellow}{\hspace{0.2em}#1 }
        \fi
    \else
        \pgfmathsetmacro{\PercentColor}{max(min(100.0*(\MidNumber - #1)/(\MidNumber-\MinNumber),100.0),0.00)} %
        \ifdim #1 pt > \Ten pt
            \hspace{-0.33em}\colorbox{green!\PercentColor!yellow}{#1}
        \else
            \hspace{-0.33em}\colorbox{green!\PercentColor!yellow}{\hspace{0.2em}#1 }
        \fi
    \fi
  }}

\newcolumntype{R}{>{\collectcell\ApplyGradient}c<{\endcollectcell}}
\usepackage[colorinlistoftodos,prependcaption]{todonotes} 
\usepackage{xargs}
\newcommandx{\unsure}[2][1=]{\todo[inline,linecolor=blue,backgroundcolor=red!25,bordercolor=red,#1]{#2}}
\newcolumntype{L}[1]{>{\raggedright\let\newline\\\arraybackslash\hspace{0pt}}m{#1}}

\journal{Journal of Commodity Markets}

\begin{document}
	
\begin{frontmatter}

\title{
	Short- and long-term forecasting of electricity prices using embedding of calendar information in neural networks}


\author[add1,add2]{Andreas Wagner\corref{cor1}}
\author[add1]{Enislay Ramentol}
\author[add1]{Florian Schirra}
\author[add2]{Hendrik Michaeli}
\address[add1]{Fraunhofer Institute for Industrial Mathematics ITWM, Department for Financial Mathematics, Fraunhofer-Platz 1, 67663 Kaiserslautern, Germany}
\address[add2]{Karlsruhe University of Applied Sciences, Moltkestraße 30, 76133 Karlsruhe, Germany}
\cortext[cor1]{andreas.wagner@h-ka.de}


\begin{abstract}
Electricity prices strongly depend on seasonality of different time scales, therefore any forecasting of electricity prices has to account for it.
Neural networks have proven successful in short-term price-forecasting, but complicated architectures like LSTM are used to integrate the seasonal behaviour.
This paper shows that simple neural network architectures like DNNs with an embedding layer for seasonality information can generate a competitive \revTwoOut{but also superior }forecast.
The embedding-based processing of calendar information additionally opens up new applications for neural networks in electricity trading, such as the generation of price forward curves.
Besides the theoretical foundation, this paper also provides an empirical multi-year study on the German electricity market for both applications and derives economical insights from the embedding layer.
The study shows that in short-term price-forecasting the mean absolute error of the proposed neural networks with \er{an }embedding layer is \revTwo{better than the LSTM and time-series benchmark models and even slightly better as our best benchmark model with a sophisticated hyperparameter optimization}.\revTwoOut{only about half of the mean absolute forecast error of state-of-the-art LSTM approaches.}
\revTwoOut{The superiority of the proposed approach is }\revTwo{The results are }supported by a statistical analysis using Friedman and Holm's tests.
\end{abstract}

%
\begin{highlights}
	\item \revTwoOut{Significant}Improvement of machine-learning-based short-term electricity price forecasting using simpler models than current state-of-the-art
	\item Application of machine-learning to the generation of hourly price forward curves (HPFC)
	\item \revTwo{Case-study on the German electricity market comparing the proposed approach and different benchmarks from the literature including a statistical analysis}
\end{highlights}

\begin{keyword}
	Machine Learning\sep Neural Networks\sep Embedding\sep Electricity Market\sep Spot Price\sep Forecasting\sep Price-Forward Curve\sep Renewables
\end{keyword}

\end{frontmatter}

\newpage
\newpage

\section{Introduction}\label{sec:introduction}
\begin{figure}[tb]
\caption{Hourly EPEX day-ahead electricity prices for Germany on different time scales}
\includegraphics[viewport=0bp 120bp 842bp 550bp,scale=0.65]{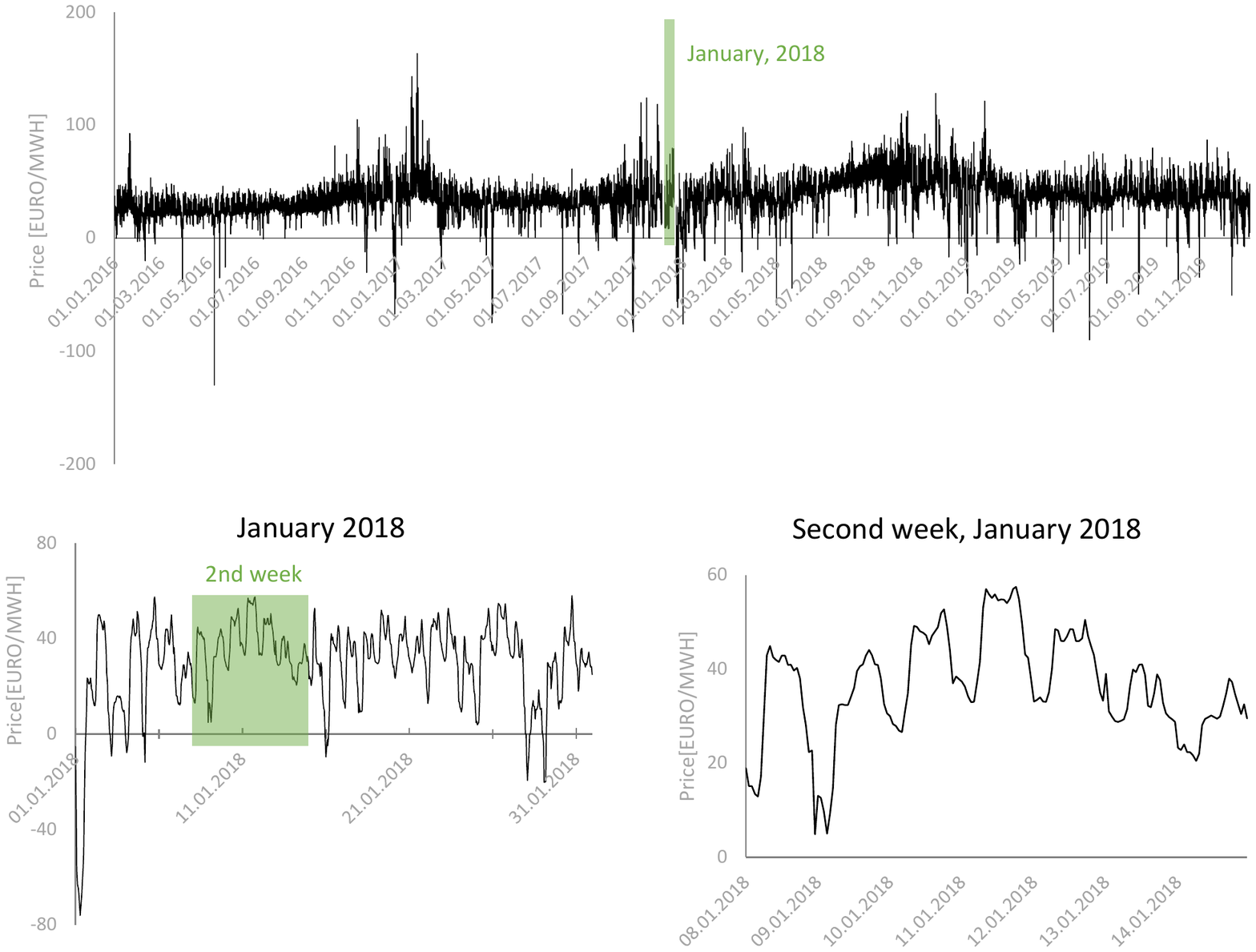} 
	\label{fig:prices}
\end{figure}

Forecasting electricity prices is an important task in the trading process of energy utilities.
This paper focuses on short-term price forecasting in the day-ahead market and long-term forecasting of the price profile.
Typical applications for short-term forecasting are proprietary trading and short-term dispatching of power plants.
The long-term price profile, on the other hand, is needed to generate (usually hourly) price-forward curves from observed futures market prices.
Both forecasting tasks face a very strong dependency on calendar information, i.e. season (as a proxy for expected temperature levels), day of the week, and hour.
This is visible in figure~\ref{fig:prices}, which shows a typical time series of EPEX day-ahead electricity prices for Germany.
This paper shows how calendar information can be used in neural-network-based forecasting models to significantly improve the forecasting quality compared to the default of using dummy variables.
Our main contribution is to prove the use of embeddings as a very successful way to represent calendar information.

Our paper focuses on the following research questions:
\begin{itemize}
	\item How can calendar information be included in neural-network-based price forecasting?
	\item How does an approach based on embeddings perform in short-term price forecasting and long-term profile forecasting?	
	\item Which economic insights can be gained from the embedding layer?
\end{itemize}

We base our empirical work on the EPEX German day-ahead electricity market and use only publicly available data.

The following section~\ref{sec:related_methods} gives a literature review on electricity price forecasting (EPF), followed by section~\ref{sec:neuralNetworksAndEmbeddings} which introduces our approach based on neural networks and in particular embeddings, as well as their application to calendar information.
Empirical results are presented in section~\ref{sec:empiricalStuddy}.
Embeddings can be used to check the plausibility of the results, which is presented in section~\ref{sec:explainableAI}.
Finally section~\ref{sec:concluion} concludes.

\section{Applications and related literature on electricity price forecasting }\label{sec:related_methods}

This paper distinguishes \er{between }two applications of EPF:
\begin{itemize}
	\item Short-term price forecasting: the aim is to forecast the (usually hourly) prices for the next day as closely as possible
	\item Long-term profile forecasting: the aim is to generate a time series of hourly prices for several years into the future; in this application, the relation between the prices should be as realistic as possible (e.g. the relative behaviour of prices on a Sunday compared to prices on a weekday).
\end{itemize}

We provide more details on the applications and existing literature for both cases.

\subsection*{Short-term price forecasting}

In short-term forecasting, we want to predict the 24 hourly prices of the next day, one day-ahead.
A good forecast is needed for trading in the day-ahead market and in decision support concerning power-plant dispatch, the scheduling of an industrial plant, or trading in alternative markets like secondary control or other auxiliary services.
The prediction of electricity prices has been widely studied by the research community in areas such as financial mathematics and machine learning.

\paragraph{Overview}
\cite{aggarwal2009} gives an early overview including 47 papers published between 1997 and 2006, with topics ranging from game-theoretic to time series and machine learning models.
\cite{Weron2014} provides an extensive overview including game-theoretic, fundamental, reduced-form, statistical, and machine learning models.

One distinguishes univariate models (same model for each hour) and multivariate models (separate models for each hour) and \cite{ziel2018} show that there is no clear preference in empirical results.
The modelling methods range from time-series approaches as in \cite{Ugurlu2018, Narajewski2019}, dynamic regression and transfer functions (\cite{nogales2002}), wavelet transformation followed by an ARIMA model (\cite{conejo2005}) and weighted nearest neighbor techniques (\cite{troncoso2007}).

\paragraph{Machine learning}
There are many applications of machine learning methods in electricity price forecasting.
\cite{amjady2006} compares the performance of a fuzzy neural network with one hidden layer to ARIMA, wavelet-ARIMA, multilayer perceptron, and radial basis function network models for the Spanish market.
\cite{chen2012} use a neural network with one hidden layer on Australian data.
On the same market, \cite{mosbah2016} use a multilayer neural network focusing on forecasting the next month, but their analysis is based on data from 2005 only.
As in this study, \cite{keles2016} use neural networks to forecast prices on the EPEX German/Austrian power market and show that the machine-learning approach performs better than a competitive time-series model like seasonal ARIMA.
In recent years, following the rapid progress of deep learning, more sophisticated variants of neural networks have become popular in EPF  \citep{Lago2018,Zhu2018,Brusaferri2019,Kuo2018,marcjasz2018}.
\aw{\cite{Lago2018} compare different neural networks and show, using a Diebold-Mariano test, that deep feed-forward, GRU (gated recurrent unit) and LSTM (long-short-term memory) networks perform best on Belgian market data.
Note that the \textit{Diebold-Mariano test} refers to the test introduced in \cite{Dieboldmariano1995}. }
\revTwo{\cite{LAGO2021} also propose machine-learning models for EPF (which we will use later in our study) and introduce a methodolody to benchmark EPF-models.}
\revTwoOut{In fact, the LSTM approach tends to be the most competitive neural network setup in EPF, which can be ascertained from various recent studies (see below after the details on LSTM neural networks).
Therefore we explain LSTM in more detail and also use it as the benchmark in our study}
\revTwo{In the literature there is evidence that a LSTM approach tends to be a competitive neural network setup in EPF, which can be ascertained from various recent studies (see below after the details on LSTM neural networks).
Therefore we explain LSTM in more detail and also add it as a benchmark in our study.}

\begin{figure}
	\centering
	\caption{Internal operation of the LSTM cell}
	\label{fig:LSTM-cell}
		\includegraphics[viewport=50bp 560bp 700bp 800bp,clip,scale=0.8]{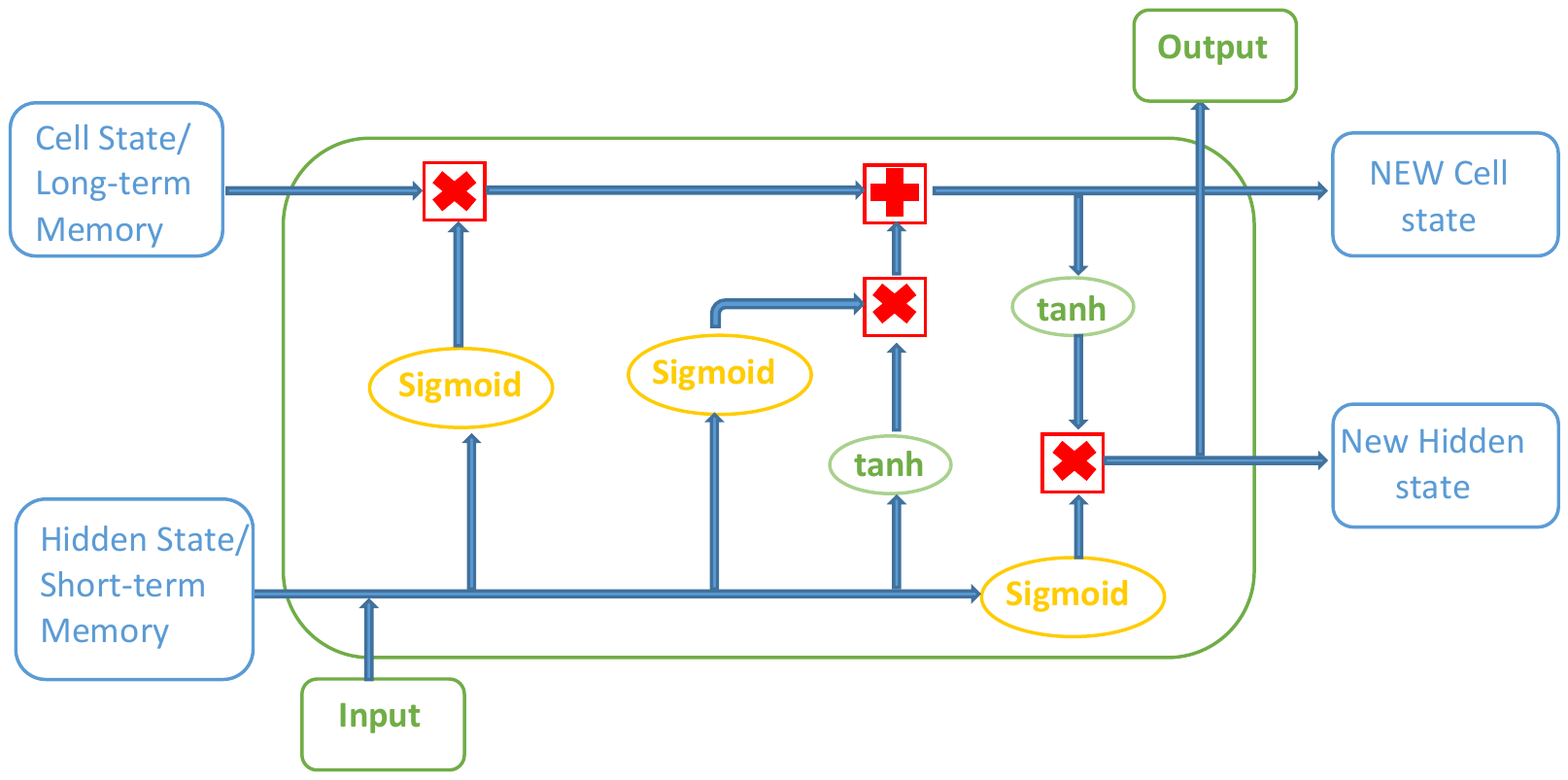}
\end{figure}

\paragraph{LSTM in EPF}
The \textbf{Long short-term memory (LSTM)}, proposed in \cite{Hochreiter1997}, is a deep learning framework that has proven successful with time series problems due to its versatility and great efficiency at \textit{remembering} information from the time-series history in the long and short term.
LSTMs are a particular variant of Recurrent Neural Networks (RNN).
The LSTM networks solved the \textit{short-term memory problem} suffered by their predecessors, RNNs.
For each time step, the LSTM cell takes three different inputs: the current input data, the short-term memory from the previous cell, and the long-term memory.
The short-term memory is also known as the hidden state, and long-term memory is generally referred to as the cell state.
Figure \ref{fig:LSTM-cell} shows the internal architecture of a LSTM cell.
Another relevant characteristic of the LSTM cell is the use of gates to regulate the information to be kept or discarded at each time step. These gates are known as the Input Gate, the Forget Gate, and the Output Gate.

In the following, we summarize the literature on the application of LSTM in EPF.
\cite{Zhu2018} use hourly price data from the New England and PJM 
day-ahead market, and train the model with different input lengths, forecasting horizons and data sizes. The experimental study shows good results in comparison with Support Vector Machines (SVM) and Decision Trees (DT).
The LSTM network designed in \cite{Zhu2018} uses the previous prices in a certain time window of length $L$ (lookback window, $L \in [4,8,12,24]$) as features.
\revTwoOut{The output is the price at time $t$.}
The authors achieved the best results using $L=24$, which shows that the price\revTwoOut{ at time $t$} depends heavily on all \revTwoOut{the }prices of the previous 24 hours.
\revTwoOut{In }\cite{Lian2018}\revTwoOut{ the authors} present a study for day-ahead electricity price forecasting using LSTM on the Australian market in the Victoria region and the Singapore market.
They use not only historical prices as features in the model but also external variables, like holidays, day of the week, hour of the day, weather conditions, oil prices and demand.
Their LSTM model predicts only the next hour, so the whole day (24 hours) is forecasted in a recursive manner.
In \cite{Bano2020} the authors study the New York Independent System Operator (NY-ISO)\footnote{NY-ISO provides electricity to different countries like United States, Canada and Israel.}.
They compare several machine-learning-techniques using one year of hourly data (2016-2017).
In their experimental study, the authors combine Multilayer Perceptrons (MLP), LSTMs, SVMs and  Logistic Regression (LR) models with two feature selection methods. They conclude that LSTM networks perform better than MLPs for the EPF problem.
A novel approach based on Gated Recurrent Units (GRU) is introduced in \cite{Ugurlu2018} to EPF in the Turkish day-ahead market.
The authors compare their proposal with seven methods based on neural networks, including the well-known LSTM and the Convolutional Neural Network (CNN).
They use a rolling window of three historical years to predict one day ahead.
In their experimental study, the\revTwoOut{ introduced} LSTM approach outperforms all selected state-of-the-art methods with a Mean Absolute Forecast Error (MAE) of 5.36.

\revTwoOut{The summary shows that LSTM is the most promising machine learning approach in EPF and it performs well on electricity markets around the world.
We will therefore use popular LSTM models as a benchmark in our study and show that our approach outperforms them, while additionally using a much simpler neural network architecture.}

\revTwo{As shown in our literature review we expect being LSTMs the most promising machine learning approach in EPF, while they perform very well on electricity markets around the world.
	However, according to \cite{LAGO2021}, it is hardly possible to define a generally advantageous state-of-the-art EPF method.
They claim that there is no sufficient evidence for LSTMs to be more accurate than other methods. 
To facilitate the comparison between different EPF approaches they also introduce two very well studied, accurate and open source benchmark models (see section below).
We take both models from \cite{LAGO2021} as benchmarks.}

\begin{figure}[tb]
\caption{Influence of wind and photovoltaic infeed on prices shown for examples in June 2018 and April 2019.}\label{fig:windsolarinfluence}
	\includegraphics[width=.49\textwidth,viewport=70bp 450bp 530bp 770bp]{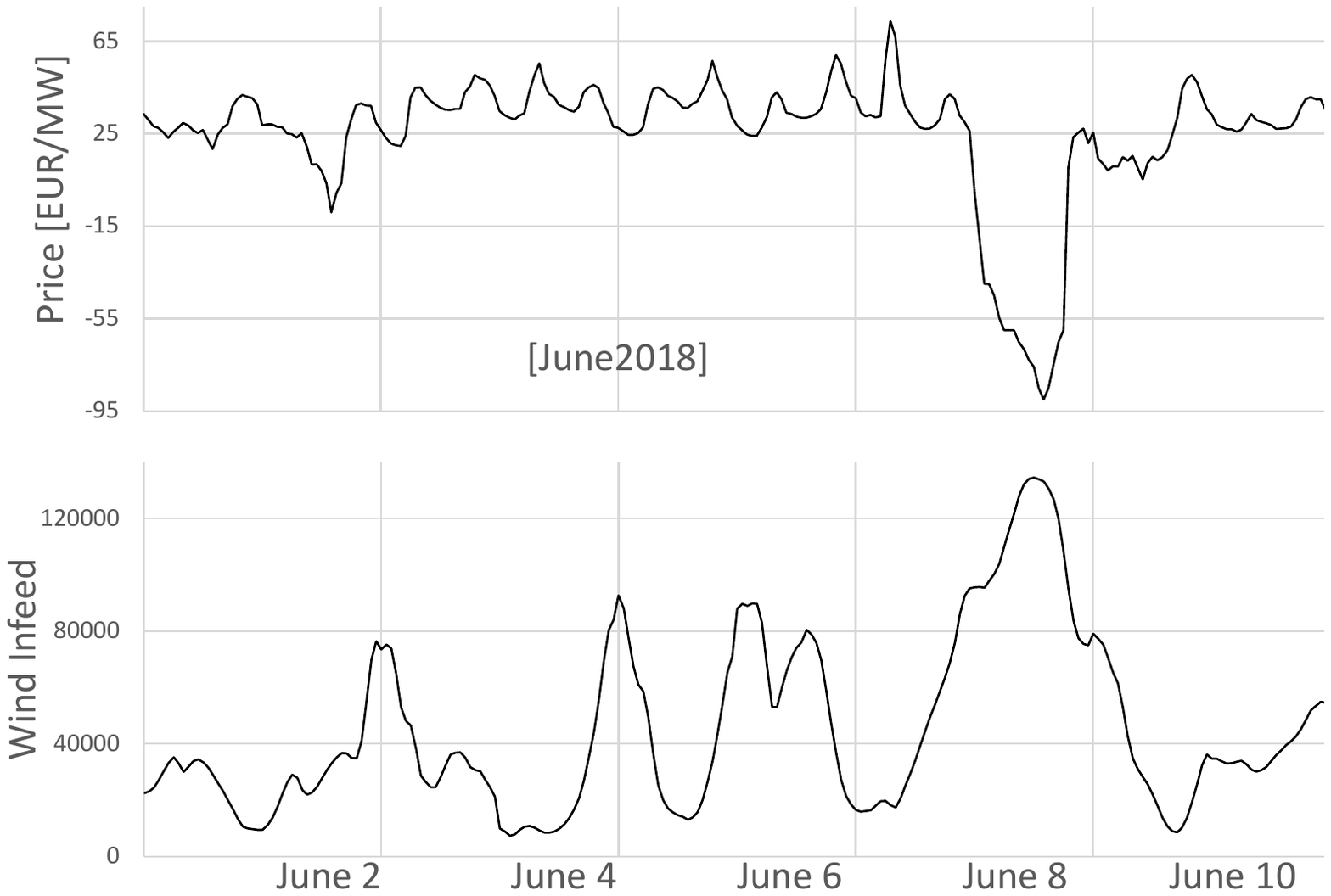}
	\includegraphics[width=.49\textwidth,viewport=70bp 450bp 530bp 770bp]{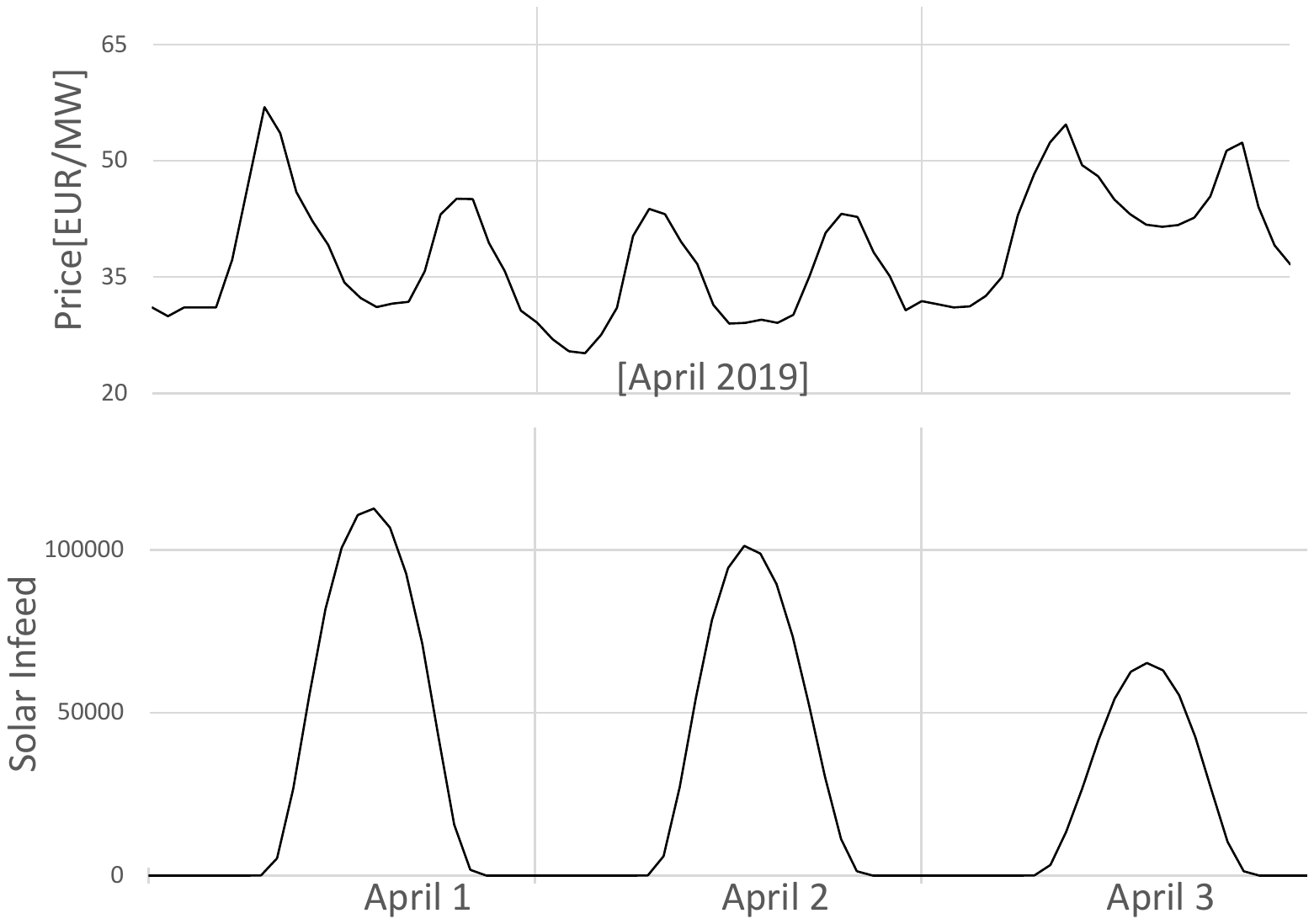}
\end{figure}

\paragraph{Features}
In the literature, the features (also dependent variables or model input) used vary from price data only \citep{Weron2005} to a whole range of fundamental data like demand, commodity prices, renewable infeed, weather data, etc. \citep{Lian2018}.
Some studies, e.g. \cite{ziel2016,schnuerchWagner2020}, use deep price information from order-book data.
To keep our study as concise as possible, we only use the most dominant features.
This is calendar information, which plays a major role in the structure of electricity prices as shown in section~\ref{sec:introduction}.
Moreover, due to the high share of renewables in the German market, we also use forecasted infeed from wind and photovoltaic (solar radiation) in some of our short-term models.
The important role of wind and photovoltaic in the German market has been proven in many studies.
The first work that recognized the need to integrate wind and photovoltaic power into models of the German power market is \cite{wagner2014}, which also proved their strong impact on the day-ahead electricity price.
Using multivariate regression methods, various authors have quantified the influence renewable infeed has on price (\cite{cludius2014,wuerzburg2013}).
Due to the regulation, higher renewable infeed generally leads to lower market prices.
This relation is shown for two exemplary periods in figure~\ref{fig:windsolarinfluence}.

\subsection*{Long-term profile forecasting}
The second application to which we apply our proposed forecasting approach is the long-term profile forecasting.
The need for a long-term profile forecast \erout{comes from}\er{is due to} the fact that for weeks, months, or even years ahead, no hourly price information can be obtained from the market until the day before delivery.
However, for many applications in electricity trading,  future price expectations are needed on an hourly granularity.
Typical applications include power plant dispatch (hourly price expectations serve as an input for deterministic or stochastic optimization), pricing of contracts for delivery like full-service contracts, production planning in industry, or even as \revTwo{a seasonality component} in stochastic price models, as in \cite{HinWa2020}, for use in risk-management or the valuation of contracts of delivery with options.

The futures, as traded on electricity markets, always have a delivery period of years, months, or other intervals.
The market participants break them down into hourly prices using historically observed day-ahead market prices (which are in hourly granularity).
The resulting price curve is called \emph{(hourly) price forward curve (HPFC)}.
A good overview of the concept of HPFC's and multiple approaches for their construction is given in \cite{Sae2017}.
In general, the generation of an HPFC can be separated into two steps:
\begin{description}
	\item[Step 1] Long-Term Profile: Generation of an hourly profile using historical hourly prices
	\item[Step 2] Absence of Arbitrage: Transformation of the curve according to market quotes for futures to \er{achieve} an arbitrage-free HPFC
\end{description}
The larger part of the existing literature is on the absence of arbitrage, which often focuses on methods to smoothen the curve.
Two prominent approaches to transform the long-term profile \erout{in an}\er{into an } HPFC are proposed in \cite{FleLem2003} and \cite{BenEtAl2007}.
\cite{FleLem2003} calculate the values of the HPFC directly by minimizing the distance to the long-term profile simultaneously to optimize the smoothness of the HPFC.
The absence of arbitrage is ensured by constraints in the optimization.
In \cite{BenEtAl2007}, on the other hand, the HPFC is not directly constructed.
Instead they calculate a correction term consisting of multiple polynomial splines.
This \erout{yields a smooth} \er{provides a smoothing } function which adds up with the long-term profile to an HPFC. Especially for the latter method there are multiple extensions.
In \cite{Sae2017} the polynomial splines are substituted by trigonometric splines. \cite{CalEtAl2007} introduces a second correction term so that there is one term for base load futures and one for peak load futures.
Our model, which we outline in the upcoming sections, can be used to generate a long-term profile.
Again, we benchmark it against popular approaches from the literature, which we detail in the following.

For the generation of the long-term profile, there are two common approaches in the literature, see \cite{KieselEtAl2019} and \cite{HinWa2020}. Up to the daily granularity we use two of the approaches - namely the dummy median and the dummy sinusoidal approaches, from \cite{HinWa2020} - as benchmark models.
To model the profile of \revTwoOut{the }hourly \revTwoOut{electricity }prices we base our approach on \cite{CalEtAl2007} and \cite{Bloe2008}, which use dummy variables with different clusters for days with a similar hourly structure.

\emph{Dummy Variables} are indicator variables combined with a certain value. They are used in cases where a state is either present or not, e.g. the month of a given date \erout{is January or it isn’t. }\er{either is January or not}.
In the literature, one commonly distinguishes between four groups of dummy variables: quarters, months, day types\footnote{We use 1. Mondays, 2. Tuesdays, Wednesdays and Thursdays, 3. Fridays, 4. Saturdays, Partial Holidays and Bridge Days and 5. Sundays and Public Holidays.} and hours. They build consecutively on each other to form the long-term forecast.
 The hourly dummy variables are clustered firstly by the quarter of the considered day and then by the day type. Hence we have 20 different hourly structures.
 The formula for the dummy-variable-based forecast is then:
\begin{equation}\label{eq:dummy}
LTF(t)_{Dummy}= \sum_{i=1}^{4}D_i^q(t) c_i^q + \sum_{i=1}^{12}D_i^m(t) c_i^m + \sum_{i=1}^{5} D_i^d(t) c_i^d + \sum_{i=1}^{4}\sum_{j=1}^{5}\sum_{k=1}^{24}D_{ijk}^h(t) c_{ijk}^h
\end{equation}

\emph{Sinusodials} Since the dummy variable approach yields profiles with jumps at every time step we introduce a second approach based on trigonometric functions.
This tends to produce smoother curves in contrast to the dummy variables.
However, it has the drawback that the periodicity ignores irregularly occurring events, such as Easter.
Therefore, it is usually combined with the dummy approach such that sinusodials are used for quarterly and monthly variations, \erout{and}\er{while } the weekly and daily profiles are \er{modelled } by dummy variables. This leads to the following formula for this approach:

\begin{equation}\label{eq:sinus}
\begin{split}
LTF(t)_{Sinusoidal} = & a_0 + a_1 \sin(\frac{2\pi t}{8760}) + b_1 \cos(\frac{2\pi t}{8760}) \\
&+ \sum_{i=1}^{5} D_i^d(t) c_i^d + \sum_{i=1}^{4}\sum_{j=1}^{5}\sum_{k=1}^{24}D_{ijk}^h(t) c_{ijk}^h
\end{split}
\end{equation}

Note that for calibration the day-ahead prices are deseasonalized with the yearly median price.
This is not an issue for the application, as the expected yearly average price is observed from traded Year-Futures.
The parameters of the dummy variables are robustly calculated using the median.
For the sinusodials, the parameters are calculated via a least-squares approach.
The results are shown in section~\ref{sec:empiricalStuddy}.

\section{Embeddings for calendar information and proposed neural network}\label{sec:neuralNetworksAndEmbeddings}

\begin{figure}
	\caption{Fully connected neural network architecture\label{fig:NN example}}
	
	\includegraphics[viewport=0bp 580bp 700bp 770bp,clip]{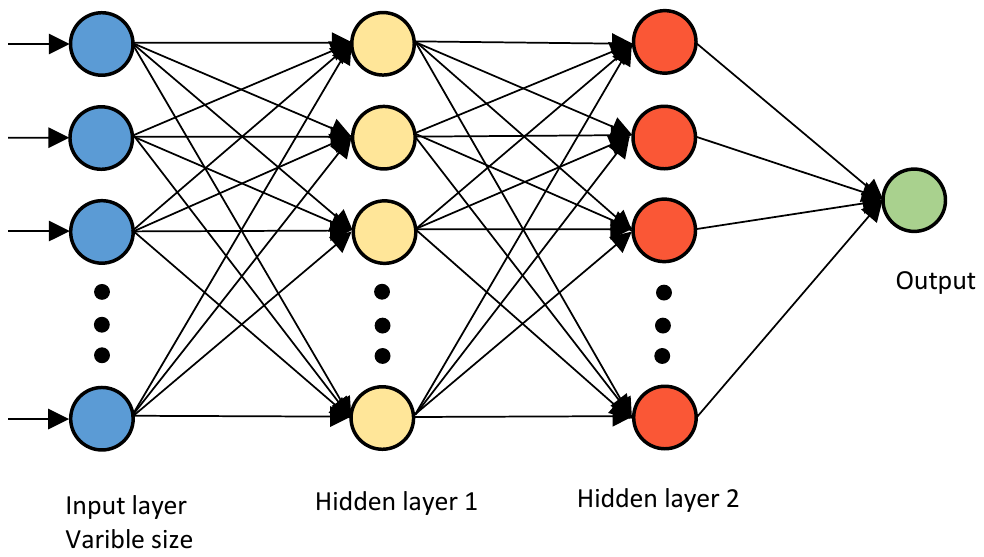}
	
\end{figure}

Dense Neural Networks (DNN) are fully connected networks.
Each neuron in a layer receives an input from all the neurons present in the previous layer as outlined in figure~\ref{fig:NN example}.
Despite the fact that this neural network architecture is quite old \citep{McCulloch1943, Hopfield1982}, they have gained great popularity in recent years due to the evolution of Deep Learning (DL). The introduction of dense layers in neural networks has brought about a considerable improvement in their performance \citep{Huang2017}.
Another contribution of DL is word embedding \citep{Bengio03aneural}, which almost 20 years after its creation has taken natural language processing (NLP) to levels never before reached.
Word embedding is one of the most fascinating areas in DL at the moment and draws the attention of a huge community of researchers \citep{Bian2014,Goyal2018, Khabiri2019}.
Algorithms like Word2vec or Doc2vec have allowed natural language analysis that, until a few years ago, was just a futuristic dream.

\begin{figure}[tb]
\center
	\caption{Example of embedding vectors generated using  Word2Vec, projected onto two dimensions using PCA. This illustrates the close proximity of related words\label{fig:Embedding_example}}.
	\includegraphics[viewport=50bp 580bp 700bp 770bp,clip]{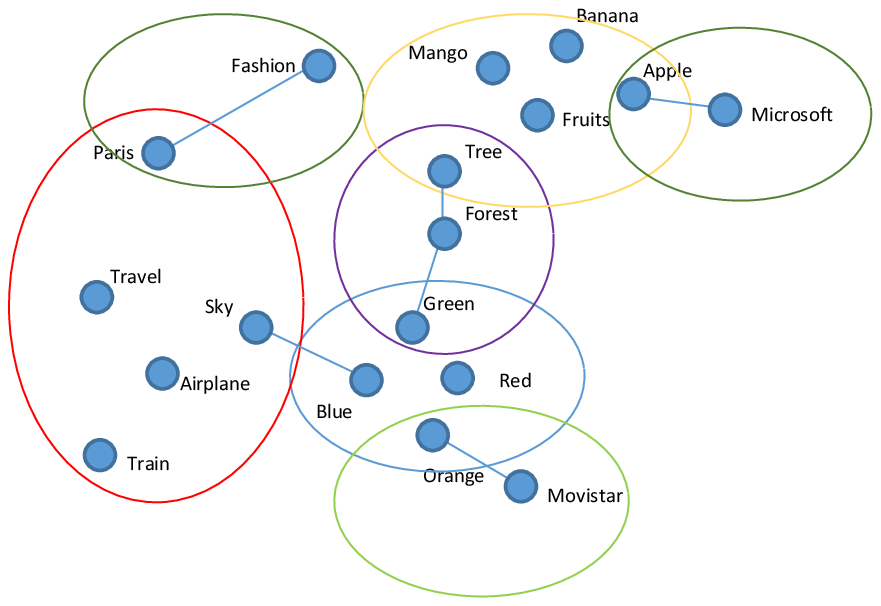}
\end{figure}

\begin{figure}[tb]
	\center
	\caption{\revTwo{Scheme of the proposed DNN with embedding layer}\revTwoOut{General scheme for a DNN with an embedding layer}\label{fig:NN_emb}
}	
	\includegraphics[scale=0.8]{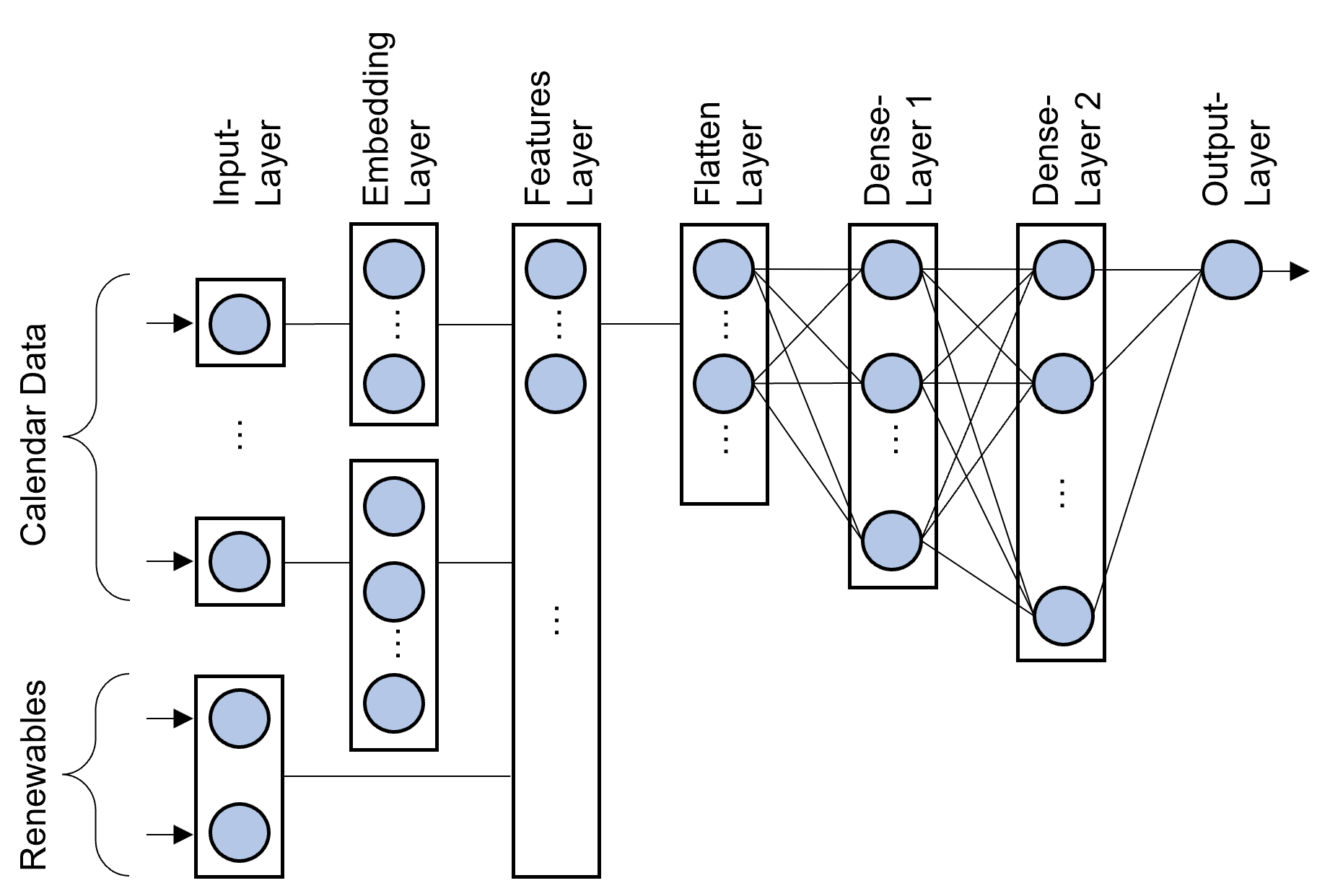}
\end{figure}

A word embedding $W$: $words\rightarrow\mathbb{R}^{n}$ is a parameterized function mapping words to high-dimensional vectors \citep{Chen2017}. Typically, the function is a lookup table, parameterized by a matrix, $\theta$, with one row for each word:
$W_{\theta}\left(w_{n}\right)=\theta_{n}$.
$W$  is initialized with a random vector for each word.
During the training, it learns meaningful vectors in order to perform some task.
The resulting embedding vectors can be interpreted as semantic features, and can be used to understand similarities or differences between words.
The distance between vectors describes their semantic similarity.
Using the numeric/semantic vectors, it is possible to \awout{do} gain additional insights, as shown in figure~\ref{fig:Embedding_example} for an NLP example.
We observe that words with a similar semantic meaning are close to each other in the 2-\aw{dimensional } projection of the embedding space.
In this paper, we utilize the advantage that an embedding layer turns categorical variables into vectors. We use the concept of word embedding to encode the calendar features (month, weekday, hour) into a neural network. The resulting embedding vectors are used for two purposes:

\begin{enumerate}
	\item As features in the neural network representing calendar information
	\item To graphically understand how electricity prices behave depending on time variables and derive economical insights
\end{enumerate}

\subsection*{Proposed neural network}\label{subsec:proposed_emb}

In this paper we propose a dense neural network with an embedding layer to encode the calendar information. Figure \ref{fig:NN_emb} \aw{illustrates the }\awout{shows a representation of our} network.
The input features are \revTwo{calendar data}\revTwoOut{embeddings} and, in some applications, infeed forecasts for renewables.
The actual neural network consists of two hidden layers with Relu (Rectified Linear Unit) activation and an output layer with linear activation (as this is a regression task).
In section~\ref{sec:empiricalStuddy} we will describe the models and parameters in more detail.

\paragraph{Embedding layer for calendar features}
The embedding layer is used to encode the calendar information.
\aw{Embeddings are an alternative to one-hot-encoding of categorical features. 
As for every layer in a neural network including the embedding layer, the number of neurons has to be chosen carefully (\textit{the number of dimensions is a hyperparameter  you  can  tweak} \cite{geron2019hands} on embeddings).
Scientific literature does not provide explicit guidelines and, to the best of our knowledge, embeddings have not been used in EPF so far.
However, there are some studies on the dimensionality of word embeddings \citep{Yin2018, Gu2020, Wendlandt2018}.
The authors point out the importance of a correct choice of the dimensionality of the embedding, as a high dimension can lead to overfitting and a very low dimension can not capture all the meaning of the categorical variable.
Most of these studies agree that the dimension must be selected empirically as it depends on the data.
We optimized the dimension between 15~\% and 35~\% in 5\% intervals and found that 25~\% gives the best overall results, even though the differences were rather small.}

We consider the following embedding variables:

\begin{itemize}
	\item Hour: The categorical variable $hour$ takes values in $\left\{ 0,1,\ldots ,23\right\}$, so in a one-hot-encoding it has dimension 24. For its embedding dimension we chose six. \erout{which follows recommendations to use 25 \% of the input dimension for the embedding space.}
	\item Weekday: The categorical variable $Weekday$ takes values in $\left\{ 0,1,\ldots , N\right\}$, where $N$ depends on the representation of holidays outlined below.
	Its embedding dimension is two.
We consider three approaches to deal with holidays:
\begin{itemize}
  \item Approach 1: weekday $\in \left\{ 0,1,\ldots ,7\right\}$ considering each weekday (Sunday, Monday, ... Saturday) separately and adding a category for holidays.
  \item Approach 2: weekday $\in \left\{ 0,1,\ldots ,9\right\}$ considering seven weekdays and three types of holidays: Partial holiday, public holiday and bridge day
  \footnote{Bridge, partial, and public holiday describe days with influence through public holidays.
   Public is the actual public holiday, partial is a public holiday in only parts of Germany and bridge describes days between a public holiday and weekends.}.
  \item Approach 3: This approach uses two embedding variables, i.e. treating holiday and weekday separately. This is motivated by the fact, that every holiday also has an associated weekday (i.e. Easter Sunday is a holiday, but also a Sunday): weekday $ \in \left\{ 0,1,\ldots ,6\right\}$ and type\_holiday $\in \left\{ 0,1,2,3\right\}$.
\end{itemize}
A list of holidays used is in appendix \ref{app:holidays}.
In this paper we present the results for Approach 2, as it gave the mean absolute deviation.
The interested reader can get detailed results for Approaches 1 and 3 by contacting the authors.

	\item Month: The categorical variable $Month$ takes values in $\left\{ 1,2,\ldots ,12\right\} $ and its embedding dimension is three.
    \item Year: The categorical variable $Year$ takes values in $\left\{ 2010,2011,\ldots  ,2019\right\} $ and  its embedding dimension is three.
    \item Cross-feature month-hour: Due to differences in daylight hours, there is a relationship in electricity demand between month and hour-of-the-day.
    The categorical variable $Month-hour$ takes values in $\left\{ 1,2,\ldots ,12*24\right\} $ and its embedding dimension is ten.
    \item Cross-feature weekday-hour: Due to differences in human behaviour (e.g. people get up later on the weekend) there is a relationship in electricity demand between weekday and hour.
    The categorical variable $weekday-hour$ takes values in $\left\{ 0,1,\ldots ,N*24\right\} $ and its embedding dimension is fifteen.
\end{itemize}

\section{Empirical performance on the German electricity market}\label{sec:empiricalStuddy}

In this section we carry out an experimental study on the German electricity market.
We distinguish the two applications as outlined in detail in section~\ref{sec:related_methods}.
In both applications we conduct a statistical analysis to show the significance of the results.
\begin{description}
  \item[Short-term forecasting:]
  For the short-term forecasting we use a DNN with an embedding layer to encode the calendar information and \erout{, as a benchmark, five different LSTM models}\er{six different benchmark models}.
  \erout{The LSTM benchmark models are recommended in the literature or have been used in similar applications \cite{Lian2018,Zhu2018}}
  \er{\revTwoOut{Three}Four of them are neural network models, \revTwoOut{two are}one is a time series approach and one is a naive method.}
  In addition to calendar information we also use forecasts on renewable infeed (wind and photovoltaic) as features \revTwo{(see below for details)}.
  
\revTwo{
We follow a training framework with daily recalibration, so we retrain the model every day with historical data available.
As we forecast the day-ahead market, which is traded at 12 o'clock for the next day, we can rely our model on prices up to the current day (which was traded the day-before) and forecasts on renewable infeed for the next day.
In our experimental study we use a history of five years to train the model.
In other words, using a five-year history of the data available up to the time the prediction model is run, the model predicts the prices for each of the next 24 hours. 
}
\revTwoOut{As in our experimental study we work with historical data, training our models with the last 5 years of data until day $X$ at 23:00, then we predict the 24 prices for day $X + 1$. This process is repeated for each day from January 1, 2015 to December 31, 2019, where a new model is trained for each day to be predicted.}

\item [Long-term profile forecasting:]
  For the long-term forecasting we use a DNN with an embedding layer and, as a benchmark, popular methods from the literature (dummy variables and sinusoidal).
  We forecast four years ahead.
   \end{description}

\paragraph{Data}

For our\revTwoOut{ experimental} study we use data from the German\footnote{Note that the German spot market had been a larger market including Austria (EPEX DE/AT) until October \nth{1} 2018.} Day-Ahead electricity market (EPEX DE) from 2010 through 2019\footnote{Due to missing data from EEX Transparency, we excluded 11/01/2010 and 10/02/2010 from our analysis.}, as traded on EPEX Spot\footnote{\url{https://www.epexspot.com/en/market-data}}.
We also use data on the expected generation from renewables in Germany, which we collect from the EEX transparency platform\footnote{\url{https://www.eex-transparency.com/power/de/production/usage/}}.
We compiled our data sets from FTP-files which we licensed from the German Energy Exchange EEX\footnote{\url{https://www.eex.com/de/marktdaten/strom}}, but the data can be viewed without a license on the corresponding websites.
More details on the data are in table \ref{tab:Data description}.
We use an Ex-Ante timestamp (i.e. hour 1 describes the price or renewable infeed for the time between 1:00 AM and 2:00 AM).

It is a known fact that neural networks work much better when variables are normalized. That is why in our experimental study the renewable variables are scaled using a technique known as standard scaler:
\begin{equation}
X_{normalized}=\frac{X-X_{mean}}{X_{stddev}}\label{eq:Scaler}
\end{equation}

The values $X_{mean}$ and $X_{stddev}$ are computed over the \revTwo{training set only}\revTwoOut{complete dataset}.
The prices were not normalized in our models because they are the output feature (makes no difference in training), except for the LSTM models, which use prices also as an input feature.

\begin{table}[tbp]
\caption{Data description}
\label{tab:Data description}
\centering
\begin{tabular}{lccc}
	\toprule	
\multicolumn{1}{c}{\multirow{2}{*}{}} & \multirow{2}{*}{\textbf{EPEX DE(/AT)}} & \multicolumn{2}{c}{\textbf{Expected production}}\tabularnewline
\cmidrule{3-4}
 &  & \textbf{Photovoltaic} & \textbf{Wind}\tabularnewline
\midrule
\textbf{Features} & {date, price} & {date, expected volume} & {date, expected volume}\tabularnewline
\textbf{Start date} & {1/1/2010} & {1/1/2010} & {1/1/2010}\tabularnewline
\textbf{Final date} & {31/12/2019} & 31/12/2019 & {31/12/2019}\tabularnewline
\bottomrule
\end{tabular}

\end{table}

The evaluation metric used in this study is the mean absolute error (MAE):
\begin{equation} 
MAE=\frac{1}{H}\overset{H}{\underset{i=1}{\sum}}\left|P_{r}-P_{p}\right|\label{eq:MAE} ,
\end{equation}
where $H$ is the number of hours, $P_{r}$ is the realized price on the exchange and $P_{p}$ is the predicted price.



\subsection*{Short-term forecasting}
This section summarizes the results of the day-ahead forecasting of hourly prices.
We compare our proposed neural networks using embeddings and benchmarks from the literature.

\paragraph{Setup}
For the experimental study we design different configurations for\revTwoOut{ the} DNN and LSTM models, see table \ref{tab:DNN parameters} and \ref{tab:lstm_parameters}.

\aw{In general there are alternative approaches to include calendar information in neural networks. 
We compare the embedding approach to two alternatives:
\begin{itemize}
\item Using ordinal variables:
\begin{itemize}
    \item weekday: numeric variable $\{0..9\}$
    \item month:  numeric variable $\{1..12\}$
    \item hour: numeric variable $\{0..23\}$
\end{itemize}
\item Using a periodicity function for month and hour. To take into account the periodicity of months and hours we project these values onto a circle and use the two-dimensional projections as features as in \citep{schnuerchWagner2020}.
\begin{itemize}
    \item weekday: numeric variable $\{0..9\}$
    \item month: see equ.~\eqref{eq:month}
    \item hour: see equ.~\eqref{eq:hour}
\end{itemize}
\end{itemize}
}

\fs{
\begin{equation}
month_{x}\left(t\right):=\sin\left(\frac{2\pi m}{12}\right),month_{y}\left(t\right):=\cos\left(\frac{2\pi m}{12}\right)\label{eq:month}
\end{equation}
}

\fs{
\begin{equation}
hour_{x}\left(t\right):=\sin\left(\frac{2\pi m}{24}\right),hour_{y}\left(t\right):=\cos\left(\frac{2\pi m}{24}\right)\label{eq:hour}
\end{equation}
}

To select the configurations for the DNN\revTwoOut{ and the LSTM models} we follow recommendations from \cite{Kapoor2019} and use the formula in equ.~\eqref{eq:neurons} (as also referenced in a blog-post \citep{Hyperparater2018} on  the  \textit{towardsdatascience.com}-website, a platform very popular among practitioners):
\begin{equation}
N_{h}=\frac{N_{s}}{\left(\alpha*\left(N_{i}+N_{o}\right)\right)}\label{eq:neurons}
\end{equation}

$N_{i}$ is the number of input neurons, $N_{o}$ the number of output neurons, $N_{s}$ the number of samples in the training data, and $\alpha$ represents a scaling factor that is usually between 2 and 10.
We calculate and compare the following models to forecast the 24 hourly prices of the next day.


\aw{We compare the following model architectures.
For most of them we do calculations with two sets of features, namely only calendar information as well as calendar information and renewables (forecasts on the infeed of wind and photovoltaic/solar energy).}
\begin{description}


\item[Naive] \fs{A naive model using past prices.
The output of the naive method for hour $h$ of date $d$ is the price at hour $h$ of the last observed day of the same type (e.g. working day, Saturday).
}

\item[LEAR] \revTwo{\textbf{L}ASSO \textbf{E}stimated \textbf{A}uto-\textbf{R}egressive model is one of two benchmark models presented in \cite{LAGO2021}.
The model is based on a parameter-rich ARX (\textbf{A}uto-\textbf{R}egressive with e\textbf{x}ogenous features) structure which is estimated by LASSO (least-absolute-shrinkage and selection operator).
They show that long calibration windows (three and four years) lead to the best results.
For our LEAR model we chose five years as calibration window, as we used the same time frame for the recalibration of our DNN models.}

\item[LSTM] Neural-network with LSTM-architecture and configuration as in table~\ref{tab:lstm_parameters}.






\item[DNN-Lago] \revTwo{Dense neural network, which is the second benchmark model presented in \cite{LAGO2021}.
This model is based on a multivariate framework.
The input features and hyperparameters of every model configuration are calculated in a separate pre-processing using Bayesian optimization.
We use the first five years of our dataset for the hyperparamter optimization.
The resulting configurations are shown as configurations c4 and c5 in table~\ref{tab:DNN parameters}.
}

\item[DNN-ordinal] \fs{Dense neural network with three different configurations c1, c2 and c3 as shown in table~\ref{tab:DNN parameters}. 
This model is also evaluated without calendar information using only renewables as features.
}


\item[DNN-sin-cos] \fs{Dense neural network with three different configurations c1, c2 and c3 as shown in table~\ref{tab:DNN parameters}. Calendar information is encoded as a circle function as in \cite{schnuerchWagner2020}}.



\item[DNN-embedding] {Our approach as presented in the previous section. 
We use three different configurations c1, c2 and c3 as shown in table~\ref{tab:DNN parameters}. 
}

\end{description}

The models are trained on the last five years preceding the day we forecast and are retrained daily.
\awout{
Most of the LSTM are computationally very intensive to train so we train them on a yearly interval, but for some easier configurations, we also perform a daily training for comparison with DNN.
More details will be found in the following paragraph.
}
Training the LSTM models is computationally extensive and could take more than one day in practice. For this reason we use a simple configuration for our experimental study.
\revTwo{All DNN configurations are summarized in table~\ref{tab:DNN parameters}.
	We use MSE as loss function.
	Note that we do show only the most important hyperparameters for the sake of clarity.}

\begin{table}[tbp]
\centering
\caption{Dense neural network parameters\label{tab:DNN parameters}}

\begin{tabular}{llllll}
\toprule
\textbf{Parameters} & {\textbf{c1}} & {\textbf{c2}} & {\textbf{c3}} & {\textbf{c4}} & {\textbf{c5}}\tabularnewline
\midrule
\textbf{Hidden layers} & 1 & 2 & 2 & 2 & 2\tabularnewline
\textbf{Neurons per layer} & 2085 & 128/128 & 2285/1024 & 484/381 & 234/203\tabularnewline
\textbf{Activation layers} & Relu & Relu & Relu & Sigmoid & Relu\tabularnewline
\textbf{Epochs} & 10 & 10 & 10 & auto & auto\tabularnewline
	\textbf{Optimizer}  &  RMSprop & RMSprop & RMSprop & Adam&Adam\tabularnewline
\bottomrule
\end{tabular}
\end{table}

%

\begin{table}[tbp]
	\centering
	\caption{LSTM parameters \label{tab:lstm_parameters}}
	
	\begin{tabular}{lll}
		\toprule
		\textbf{Parameters} &  \tabularnewline
		\midrule
		\textbf{Hidden layers} & 3 \tabularnewline
		\textbf{Neurons per layer}  & 10/10/24 \tabularnewline
		\textbf{Type of layer} & LSTM/LSTM/dense \tabularnewline
		\textbf{Activation layers} & Relu\tabularnewline
		\textbf{Epochs}  & 10\tabularnewline
		\textbf{Optimizer}  &  Adam \tabularnewline
		\bottomrule
	\end{tabular}

	\end{table}

%
%

\paragraph{Results}
We predict the next 24 hours using the past five years of data.
We repeat these experiments for every day starting on the \nth{1} of January  2015.
Note that, therefore, all results are out-of-sample and provide a valid  benchmark for use in practice.
In the following, we present the mean hourly absolute error per year, for every different configuration and training sample.
Any interested reader\fs{ can get detailed results (every hour) by contacting the authors}\fsout{find the detailed results (every hour) in the website associated with this paper}.
\er{Table \ref{tab:Benchmark} shows the results.}
\erout{Tables~\ref{tab:DNN_daily_recalibration} shows results for the proposed dense neural networks using embeddings for calendar information.
Table~\ref{tab:all_LSTM} shows the results for the univariate and multivariate model using LSTM, table~\ref{tab:all_LSTM-1} gives the results for benchmark models from the literature.}

\begin{table}
\centering
\caption{Mean absolute error for all benchmark methods and the proposed models\label{tab:Benchmark}}

\begin{tabular}{llcRRRRRR}
\hline 

\textbf{Method} & \textbf{Features} & \textbf{Config}. & \textbf{2015} & \textbf{2016} & \textbf{2017} & \textbf{2018} & \textbf{2019} & \textbf{all}\EndTableHeader\tabularnewline
\hline 
{LSTM} & renewables & - &  7.12 & 6.52 & 7.94 & 9.38 & 8.77 & 7.94\tabularnewline
\hline 
Naive & - & - & 7.34 & 6.19 & 9.89 & 10.43 & 9.77 & 8.72\tabularnewline
\hline 
{LEAR} & renewables & - &  4.22 & 4.26 & 4.70 & 5.92 & 4.84
& 4.79\tabularnewline
\hline 
\multirow{2}{*}{DNN-Lago} &\multirow{2}{*}{renewables} & c4 & 3.68 & 3.70 & 5.32 & 5.01 & 4.52 & 4.45\tabularnewline
 &  & c5 & 3.76 & 3.52 & 4.56 & 4.77 & 4.53 & 4.23\tabularnewline
\hline 
\multirow{9}{*}{DNN-ordinal} 
& \multirow{3}{*}{renewables} & c1 & 8.30 & 6.98 & 9.76 & 9.77 & 9.49 & 8.85\tabularnewline
 &  & c2 & 8.23 & 6.69 & 9.66 & 9.52 & 9.11 & 8.64\tabularnewline
 &  & c3 & 8.51 & 6.88 & 9.56 & 9.83 & 8.72 & 8.70\tabularnewline
\cline{2-2}
& \multirow{3}{*}{calendar} & c1 & 7.76 & 6.76 & 9.87 & 10.06 & 9.10 & 8.71\tabularnewline
 &  & c2 & 7.78 & 6.71 & 9.77 & 10.15 & 9.11 & 8.70\tabularnewline
 &  & c3 & 7.21 & 6.47 & 9.38 & 9.87 & 8.65 & 8.31\tabularnewline
\cline{2-2} 
 & \multirow{3}{*}{+ renewables} & c1 & 5.44 & 4.98 & 7.05 & 6.73 & 6.97 & 6.23\tabularnewline
 &  & c2 & 5.57 & 4.92 & 7.03 & 6.83 & 7.14 & 6.30\tabularnewline
 &  & c3 & 5.16 & 4.66 & 6.52 & 6.63 & 6.15 & 5.82\tabularnewline
\hline 
\multirow{6}{*}{DNN-sin-cos} & \multirow{3}{*}{calendar} & c1 & 7.06 & 5.93 & 8.93 & 10.05 & 8.18 & 8.03\tabularnewline
 &  & c2 & 6.71 & 5.67 & 8.75 & 9.33 & 8.01 & 7.69\tabularnewline
 &  & c3 & 6.26 & 5.47 & 8.48 & 8.70 & 7.59 & 7.30\tabularnewline
\cline{2-2} 
 & \multirow{3}{*}{+ renewables} & c1 & 4.29 & 3.93 & 5.58 & 5.89 & 5.83 & 5.10\tabularnewline
 &  & c2 & 4.13 & 3.72 & 5.43 & 5.51 & 5.39 & 4.84\tabularnewline
 &  & c3 & 3.88 & 3.49 & 4.85 & 4.90 & 4.77 & 4.38\tabularnewline
\hline 
\multirow{6}{*}{DNN-embedding} & \multirow{3}{*}{calendar} & c1 & 5.92 & 5.06 & 8.11 & 8.44 & 8.41 & 6.98\tabularnewline
 &  & c2 & 6.04 & 5.25 & 8.16 & 8.38 & 7.37 & 7.04\tabularnewline
 &  & c3 & 5.86 & 5.00 & 7.87 & 8.19 & 7.24 & 6.83\tabularnewline
\cline{2-2} 
 & \multirow{3}{*}{+ renewables} & c1 & 3.78 & 3.42 & 5.12 & 4.93  & 5.00  & 4.45 \tabularnewline
 &  & c2 & 3.82 & 3.38  & 5.09 & 4.98  & 4.78  & 4.41 \tabularnewline
 &  & c3 & 3.50 & 3.21 & 4.69 & 4.65 & 4.46 & 4.10\tabularnewline
\hline 
\end{tabular}
\end{table}

\begin{figure}[tb]
\caption{Hourly and predicted prices for September of 2016 using model DNN-emb-renew-c3.\label{fig:Sep-2016_renow}}
\includegraphics[viewport=80bp 580bp 595bp 790bp]{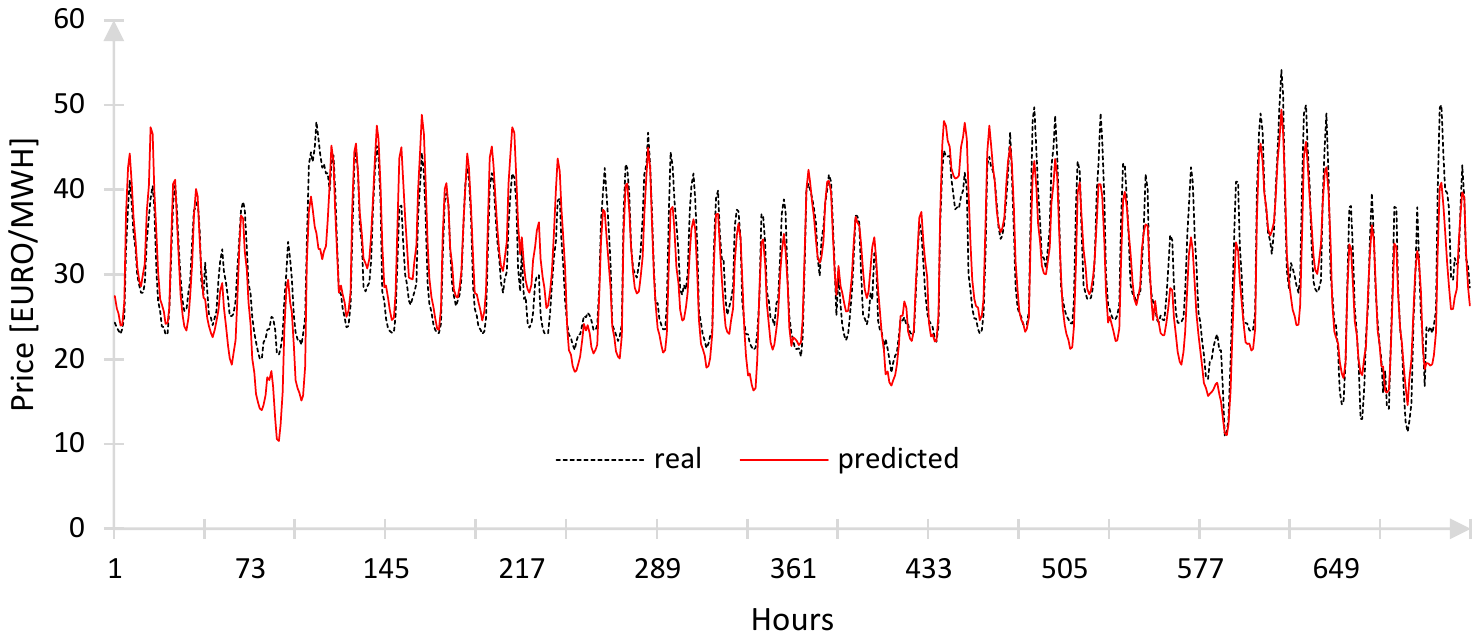}
\end{figure}

\revTwo{Our proposed approach based on embeddings performs well compared to the benchmarks.
The DNN-embeddings + renewables approach has an overall MAE of 4.10 EUR/MWh for its best configuration (DNN-emb-renew-c3). 
Other competitive approaches include DNN-Lago (4.23) and DNN-sin-cos (4.38).}
Figure \ref{fig:Sep-2016_renow} shows an example of hourly prediction for September of 2016 using renewables and embeddings for calendar information.
It can be observed that our model is able to \revTwoOut{forecast the prices with a low error (mean of $2.88$ in September) and }nicely capture the seasonal structure.

\revTwo{We can conclude that our proposed method is very competitive with the existing state-of-the-art machine-learning based forecast of electricity prices.
However, we think that it provides those good results with a fairly simple model architecture.
In order to statistically support our findings we conduct an extensive analysis using non-parametric tests in the following.
}

\subsection*{Statistical analysis}\label{subsec:statistical}
In this section we carry out a statistical analysis using non-parametric tests as recommended in \cite{demsar2006statistical} in order to compare the different algorithms and configurations appropriately.
We use the open-source software tool KEEL (Knowledge Extraction based on Evolutionary Learning) from \cite{KEEL2011}\footnote{http://www.keel.es}.
Our goal is to find out if there are significant differences between our models using embeddings and the benchmark algorithms.

First, we use Friedman’s aligned-ranks test to detect statistical differences among a set of algorithms \citep{Friedman1937}.
The Friedman test computes the average aligned-ranks of each algorithm, obtained by computing the difference between the performance of the algorithm and the mean performance of all algorithms for each dataset.
In our setting, every daily MAE can be considered a new result for a different dataset.
It holds: the lower the average rank, the better the algorithm.
If the Friedman test finds significant differences between the compared algorithms, we check if the control algorithm (the one with the smallest rank) is significantly better than the others using Holm’s posthoc test \citep{holm79}.
We use a significance level of $\alpha = 0.05$.
\fs{We are aware that the Diebold-Mariano test is a common method when comparing different forecasting approaches. However we refer to \cite{Die2013}, where the originator of the Diebold-Mariano test evaluates the usage of his test in research. He argues that it is misused in a setting where different models are compared in a pseudo out of sample analysis. We are in this setting and therefore follow the advice by using another test, i.e. the Friedman test.}
In the comparison we have 1826 samples, i.e. one per day for the five years (2015-2019) we are forecasting.
We compare a total of \fs{11 models}.
Table~\ref{tab:Friedman_test} shows the average ranks obtained by each method in the Friedman test and the adjusted $p$-values obtained by Holm’s Posthoc using \fsout{DNN-emb-renew-c3} \fs{the DNN-embddings + renewables  configuration 3 (DNN-emb-renew\_c3)} as control algorithm.
The $p$-value computed by the Friedman test is 0 \fs{or close to 0}, which means that there exist significant differences between the compared algorithms and the hypothesis of equivalence can be rejected.
We can observe that the best rank corresponds to the DNN-emb-renew\_c3 followed by 
DNN-Lago-renew-c5 and DNN-Lago-renew-c4.
From the $p$-values computed by the Holm's Posthoc test we can conclude that our approach based on embedding variables (configuration 3) is statistically superior to all the compared algorithms in table \ref{tab:Friedman_test} with the exception of DNN-Lago-renew-c5, in which case the null hypothesis cannot be rejected.

\begin{table}
\centering
\fs{
\caption{Average Friedman Ranking and adjusted $p$-values using Holm's Posthoc
procedure for the 5 years of daily predictions, using  DNN-emb-renew\_c3 as the
control algorithm}\label{tab:Friedman_test}}

\begin{tabular}{llrc}
\toprule
\textbf{i} & \textbf{Method} & \textbf{Friedman ranking} & \textbf{Adjusted p-value}\tabularnewline
\midrule
1 & DNN-emb-renew-c3 & 4.0975 & -\tabularnewline
2 & DNN-Lago-renew-c5 & 4.262 & 0.669005\tabularnewline
3 & DNN-Lago-renew-c4 & 4.592 & 0.000093\tabularnewline
4 & DNN-emb-renew-c2 & 4.825 & 0\tabularnewline
5 & DNN-emb-renew-c1 & 4.9181 & 0\tabularnewline
6 & LEAR-renew & 5.0452 & 0\tabularnewline
7 & DNN-emb-cal-c3 & 7.1679 & 0\tabularnewline
8 & DNN-emb-cal-c1 & 7.3182 & 0\tabularnewline
9 & DNN-emb-cal-c2 & 7.6054 & 0\tabularnewline
10 & Naive & 8.043 & 0\tabularnewline
11 & LSTM & 8.1257 & 0\tabularnewline
\bottomrule

\end{tabular}
\end{table}

\begin{figure}
\centering
\caption{$p$-values obtained by Holm's test. The algorithms have been ordered according to their Friedman rank. The numbers on the axes correspond to the number ($i$) in table \ref{tab:Friedman_test}. \label{fig:friedman_NxN}}

\includegraphics[viewport=0bp 0bp 432bp 288bp,scale=0.9]{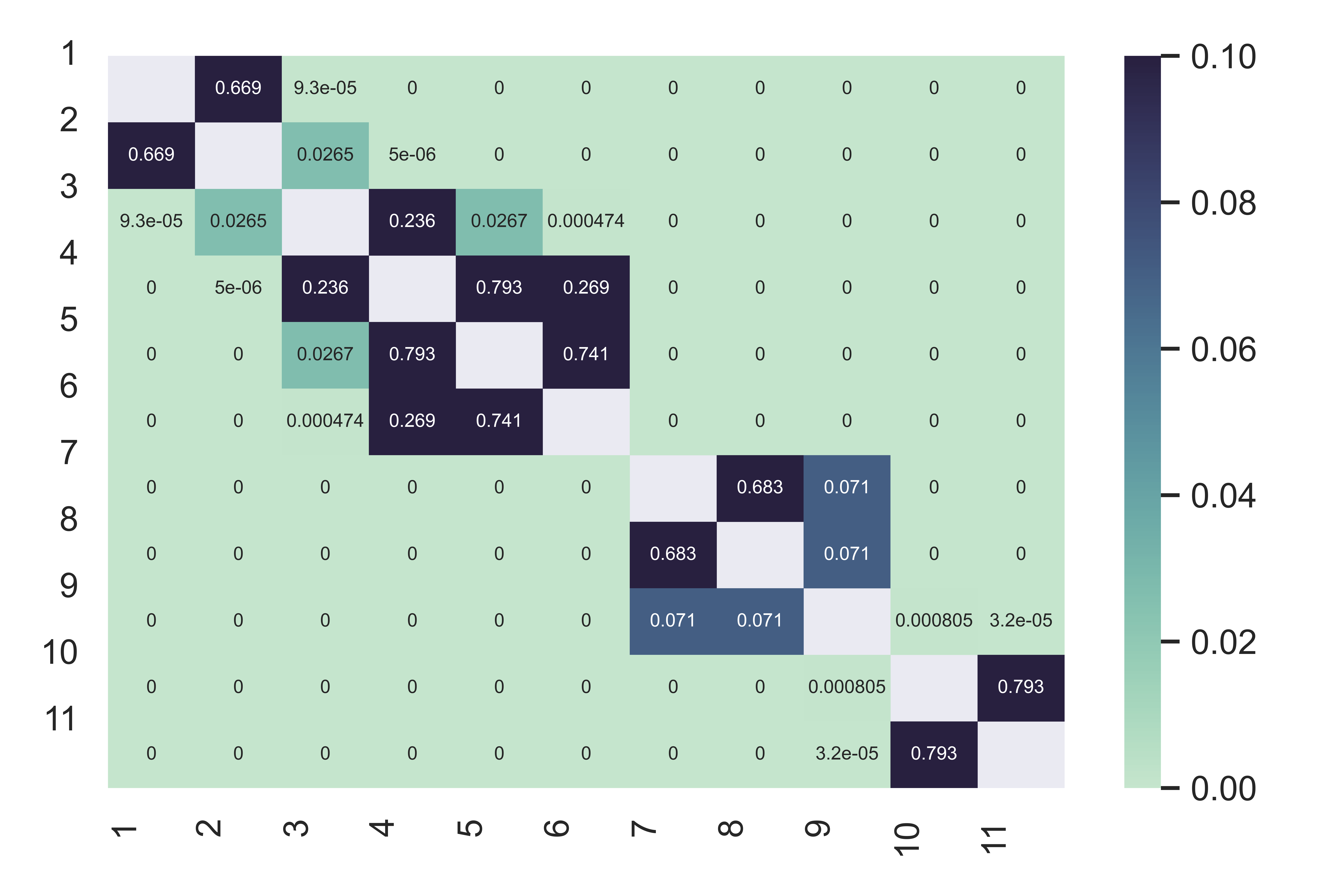}

\end{figure}

To compare all models to each other, we perform a $NxN$ Friedman’s aligned-ranks test and compute the $p$-values using the Holm's Posthoc test.
Figure \ref{fig:friedman_NxN} shows a heat map with all the $p$-values obtained with the Holm's test. The algorithms have been ordered according to the Friedman ranking. In the $x$-axis the algorithms are ordered from left to right (best to worst).
In the $y$-axis the algorithms are ordered from top to bottom (best to worst).
\revTwoOut{The darker the color, the lower the $p$-value - black color means $p$-value \er{of } zero or close to zero.
Below the diagonal, it can be seen that in all cases our models with \erout{or without} renewables dominate the state-of-the-art methods\erout{ significantly}.
This proves that our approach is statistically superior to the state-of-the-art models on our dataset.\erout{\textit{even without using renewable forecasts}}}

\begin{figure}[tb]
\centering
\caption{Boxplots of the errors for different embedding variables for DNN-emb-renew-c3. \label{fig:boxplots}}
\includegraphics[scale=0.4]{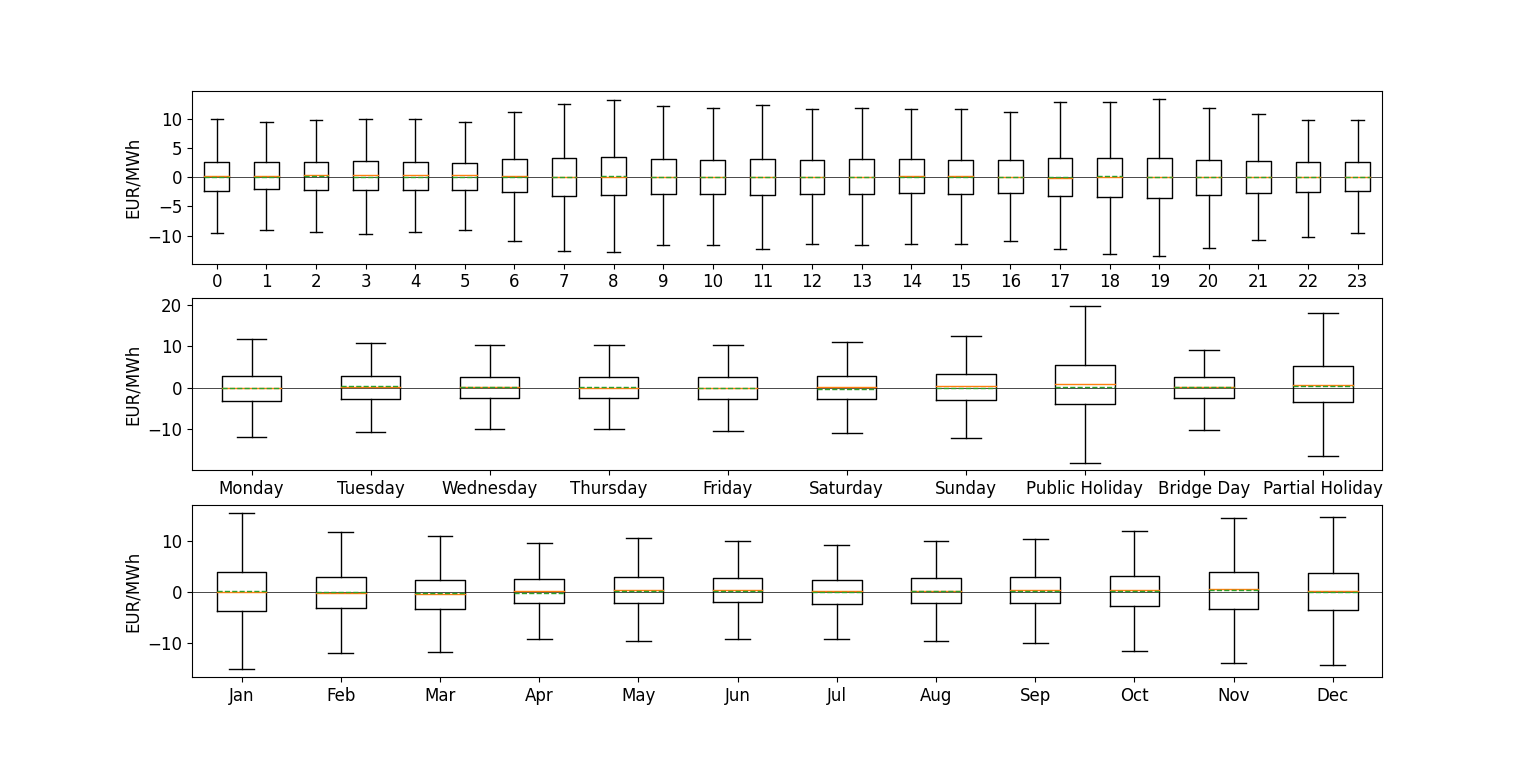}
\end{figure}

\begin{figure}[tb]
\centering
\caption{Partial autocorrelations of the absolute error of the forecast and the Day-Ahead prices. \label{fig:pacf}}
\includegraphics[scale= 0.65]{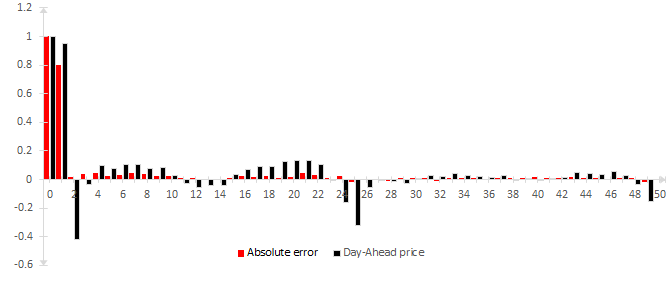}
\end{figure}

From the statistical study performed, we conclude that the use of embeddings to turn the categorical calendar variables into vectors significantly improves the predictions of electricity prices and \revTwoOut{even }allows the use of simple neural network architectures.
\fs{
\subsection*{Analysis of errors}
The claim that embeddings can be used to gain insight into the \textit{black box} neural network only holds true if there is no seasonality structure left in the errors. 
Therefore, we present an analysis of the errors and show that there is no significant pattern in the errors caused by calendar information (hour, type of day, etc.). 
We use the best model configuration, i.e. DNN-emb-renew-c3.

Boxplots of the errors for different embedding variables in Figure \ref{fig:boxplots} show that there is no particular pattern.
There are differences in the scatter range from time to time. 
In particular it is noticeable that the mean values (green dashed lines) \revTwo{are close or even to zero}.
Therefore it can be assumed that the model is unbiased or at least without significant bias and it accounts - at least in the mean - for seasonal effects. 
A comparison of the autocorrelation of absolute errors and day-ahead prices further justifies this conclusion.
\revTwo{Figure \ref{fig:pacf} shows autocorrelations for one-week (7*24 hours).
A seasonality is obvious} in day-ahead prices (as expected), whereas for the absolute errors almost all autocorrelations are below 0.05 and thus negligible. 
\revTwo{We also analyzed autocorrelation for multiple months with the same conclusions.}
In particular, the errors do not reproduce the autocorrelation structure of day-ahead prices, proving that the model indeed captures the seasonality of prices.
For these reasons, we conclude that seasonal effects are largely captured by the neural network.}

\subsection*{Long-term profile forecasting}
This section shows the use of a simple dense neural network with (embedded) calendar features in order to generate a long-term profile of expected electricity prices.
We mimic the practical application: we train the model on historical data and use it to forecast an hourly profile.
Although the models can be used to forecast multiple years (in practice the next four years may be relevant), for the presentation of results we only forecast one year ahead in order to be able to calculate out-of-sample error measures on our data.
For the evaluation, we define suitable measures on the quality of the generated hourly profile.
We use two benchmark models from the literature for comparison.

\paragraph{Setup}
We train the models on January \nth{1} 2015, 2016, 2017, 2018, 2019 and use in each case the whole past from 1/1/2010 for training.
We compare two neural networks (one with cross-features and one without) and compare them to standard approaches from the literature:

\begin{description}
	\item[LTF(1)] Dense neural network, using the calendar features month, weekday, hour, and year as described in section~\ref{sec:neuralNetworksAndEmbeddings}. The different configurations are detailed in table~\ref{tab:DNN parameters}.
	\item[LTF(2)] Same as LTF(1), but additionally using cross-features for hour and type of day as well as hour and month.
	\item[LTF(3)] Benchmark method using dummy variables as outlined in section~\ref{sec:related_methods} and equ.~\eqref{eq:dummy}  using the calendar features month, day type, hour, and year.
	\item[LTF(4)] Benchmark method using sinusoidals as outlined in section~\ref{sec:related_methods} and equ.~\eqref{eq:sinus} using the calendar features month, day type, hour, and year.	
\end{description}

In a nutshell, we evaluate four models, use three different configurations for the neural-networks and five test periods, resulting in a total of 40 experiments.
In the following we only give summaries, important results, and make conclusions.
Additional results can be \fs{obtained by contacting the authors} \fsout{found on the associated web page}.

As detailed above, the generated profiles are used to create hourly price forward curves (HPFC).
In order to generate the HPFC, the hourly profiles are shifted according to observed market prices for futures\footnote{Traded futures have delivery periods of years, quarters, months or weeks. In particular no hourly prices can be observed on the market until a day before delivery.}.
For this reason, the quality or goodness-of-fit of the profiles cannot be measured as a classical forecast error.
We have to define a suitable measure\revTwoOut{ for comparison of the models}.
The measure must rate the quality of the structure of the daily and hourly profile.
Therefore we do not compare the forecasted prices with the realized prices directly, but use the following modified time-series in order to calculate error measures like an MAE or L2-error.

We use the deviation of the daily average price to the corresponding monthly average (dDev) and the deviation of hourly price to the corresponding daily average (hDev) for the hourly seasonality.
\revTwoOut{In a formal definition we get}\revTwo{This is}
\[
hDev_d^h=  p_d^h -  \frac{1}{24 }  \sum_{h=1}^{24} p_{d}^h
\]
and
\[
dDev_d=  \frac{1}{24} \sum_{h=1}^{24} \left[ p_d^h -  \frac{1}{|m(d)| } \sum_{\bar d\in m(d) }  p_{\bar d}^h \right]
\]
where $p_d^h$ is the price at day $d$ in hour $h$ and $m(d)$ is the month corresponding to day $d$. In the following, we further use the MAE (equ.~\eqref{eq:MAE}) to analyze the forecast quality.

\begin{figure}[tbp]	
	\caption{Errors of daily deviations (LTF(2)\_c1 and LTF(4)) for training with 5
		years of real spot prices\label{fig:Daily-Deviations}}
	\includegraphics[viewport=80bp 580bp 595bp 790bp]{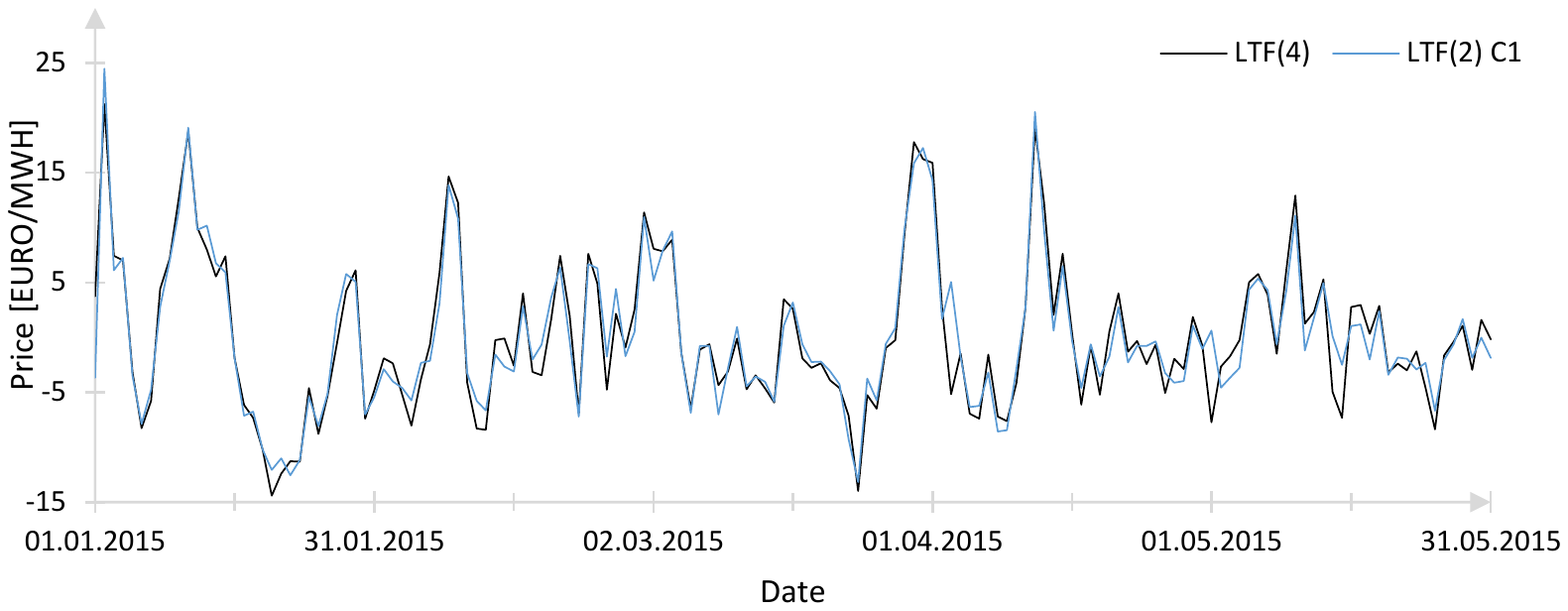}	
\end{figure}

\begin{figure}[tbp]	
	\caption{Errors of hourly deviations (LTF(2)\_c1 and LTF(4)) for training with 5
		years of real spot prices\label{fig:Hourly-Deviations}}
	\includegraphics[viewport=80bp 580bp 595bp 790bp]{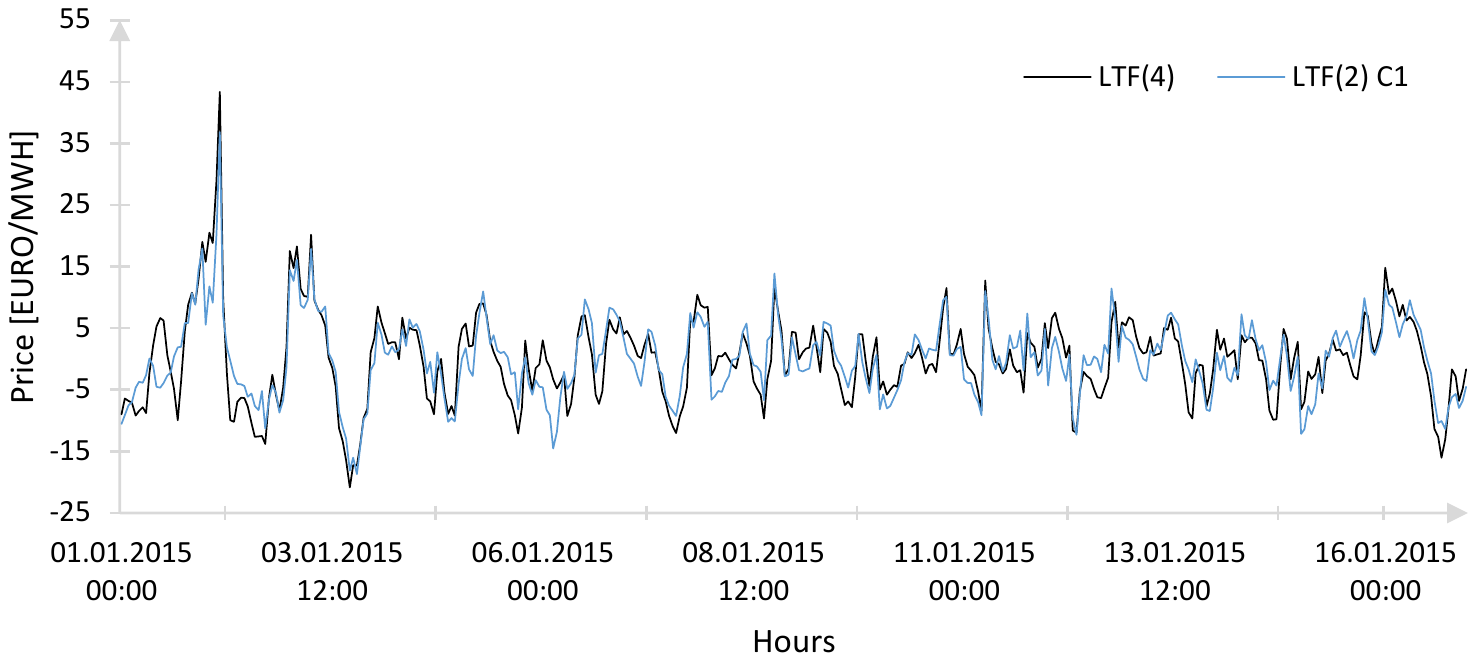}	
\end{figure}

\paragraph{Results}
\revTwoOut{As a summary we present the MAE's of the deviations $dDev$ and $hDev$, as described above. }
The forecast quality is defined as the MAE between $hDev$ ($dDev$) of the forecasted prices and $hDev$ ($hDev$) of the realized prices.
Note that for analysis we only compare the one-year-ahead forecasts, since this is the most relevant time-horizon for practitioners.
Regarding the daily deviations, we cannot find a significant difference between \revTwo{any of} the methods, see table ~\ref{tab:mae_LTF_daily}. The range of the overall MAEs is just 0.2685 or 4,71\% of the smallest overall MAE. Moreover, when comparing the best results of the benchmarks and the embedding forecasts, the difference in MAE is just 0.0077 or 0.14\% of the smallest overall MAE.
We conclude that the embedding approach is as well suited to reproduce the structure of daily prices as established methods but yields no significant improvement to the benchmark.
Concerning the hourly deviations (table~\ref{tab:mae_LTF_hourly}) we see a different picture.
In this case, the best embedding configuration - LTF(2)\_c1 - improves the overall MAE of the best benchmark method - the sinusoidal approach LTF(4)- by 0.3565 or 8.35\%. In particular, the results are strictly better for every year and do not interchange.
For a graphical overview, we reduce the number of considered methods to the best neural network configuration and the best benchmark method, i.e. LTF(2)\_c1 and LTF(4).
Figure~\ref{fig:Daily-Deviations} (figure~\ref{fig:Hourly-Deviations}) shows differences between the daily (hourly) deviations of the two methods and the real spot prices, i.e. the error of the daily (hourly) deviations.
In general, the errors follow a similar curve, which can be expected as both methods make use only of calendar information.
However, as the numbers suggest, the hourly deviations for the embedding-based neural network are smaller.

We can conclude that our neural network is applicable to the task of creating long-term profiles for forwarding curve generation.
It is even slightly better than the benchmark methods used for comparison.\revTwoOut{ but does not outperform existing state-of-the-art models as in short-term forecasting.}

\begin{table}[tbp]
	\centering
	\caption{MAE of dDev (validating \textbf{daily} seasonality). Yearly recalibration, training with data from 2010 to the year previous to the one simulated\label{tab:mae_LTF_daily}}
	
	\begin{tabular}{cRRRRRRRR}
		\toprule
		\multicolumn{1}{c}{} & \multicolumn{3}{c}{\textbf{LTF(1)}} & \multicolumn{3}{c}{\textbf{LTF(2)}} & \multicolumn{1}{c}{\textbf{LTF(3)}} & \multicolumn{1}{c}{\textbf{LTF(4)}}\tabularnewline
		\cmidrule(r{0.5em}){2-4}\cmidrule(l{0.5em}){5-7}
		\textbf{Year} & \textbf{c1} & \textbf{c2} & \textbf{c3} &\textbf{c1} & \textbf{c2} & \textbf{c3} & & \EndTableHeader \tabularnewline
		\midrule
		2015 & 4.7264 & 4.8842 & 4.8637 & 4.6813 & 4.7146 & 5.0050 & 4.8461 & 4.7952 \tabularnewline
		2016 & 4.5485 & 4.7153 & 4.7812 & 4.2544 & 4.3026 & 4.2401 & 4.6787 & 4.3661 \tabularnewline
		2017 & 7.1219 & 7.1328 & 7.3774 & 6.9513 & 7.0069 & 6.9197 & 7.0708 & 6.9616 \tabularnewline
		2018 & 6.4918 & 6.5174 & 6.5266 & 6.4808 & 6.3927 & 6.5746 & 6.4396 & 6.3725 \tabularnewline
		2019 & 6.2172 & 6.2296 & 6.3075 & 6.1461 & 6.1218 & 6.2829 & 6.1734 & 6.0571 \tabularnewline
		\midrule
		\textbf{Mean} & 5.8212 & 5.8959 & 5.9713 & 5.7028 & 5.7077 & 5.8045 & 5.8417 & 5.7105 \tabularnewline
		\bottomrule
	\end{tabular}
	
\end{table}

\begin{table}[t]
	\centering
	\caption{MAE of hDev (validating \textbf{hourly} seasonality). Yearly recalibration, training with data from 2010 to the year previous to the one simulated\label{tab:mae_LTF_hourly}}
	
	\begin{tabular}{cRRRRRRRR}
		\toprule
		\multicolumn{1}{c}{} & \multicolumn{3}{c}{\textbf{LTF(1)}} & \multicolumn{3}{c}{\textbf{LTF(2)}} & \multicolumn{1}{c}{\textbf{LTF(3)}} & \multicolumn{1}{c}{\textbf{LTF(4)}}\tabularnewline
		\cmidrule(r{0.5em}){2-4}\cmidrule(l{0.5em}){5-7}
		\textbf{Year} & \textbf{c1} & \textbf{c2} & \textbf{c3} &\textbf{c1} & \textbf{c2} & \textbf{c3} & & \EndTableHeader\tabularnewline
		\midrule
		2015 & 4.1228 & 4.5492 & 4.4911 & 4.0262 & 3.9669 & 4.0012 & 4.6098 & 4.3846 \tabularnewline
		2016 & 3.5656 & 4.2070 & 3.8957 & 3.5573 & 3.7369 & 3.3727 & 4.7161 & 3.9993 \tabularnewline
		2017 & 5.2106 & 4.6969 & 5.0940 & 4.6332 & 4.7876 & 4.7680 & 5.6534 & 5.071	 \tabularnewline
		2018 & 4.7313 & 4.8363 & 4.5289 & 4.6598 & 4.6360 & 4.6511 & 5.3895 & 4.8848 \tabularnewline
		2019 & 4.7318 & 4.8349 & 5.0043 & 4.5092 & 4.6469 & 4.5611 & 5.6929 & 4.8285 \tabularnewline
		\midrule
		\textbf{Mean} & 4.4724 & 4.6248 & 4.6028 & 4.2771 & 4.3549 & 4.2708 & 5.2123 & 4.6336 \tabularnewline
		\bottomrule
	\end{tabular}
	
\end{table}

\section{Economic insights from embedding layer}\label{sec:explainableAI}
In 2018, the new General Data Protection Regulation (GDPR) approved by the European Union entered into force, requiring citizens to have the right to an explanation regarding any algorithmic decision-making \citep{Goodman2016}. The new GDPR states that if an algorithm makes an automatic decision regarding a user, this user will have the right to obtain an explanation of how the decision was made \citep{Qureshi2017}.
In this section, we briefly show the use of the embedding vector obtained during the training of the DNN to graphically understand how the models use the calendar information in the forecast.
To visualize the resulting embedding vectors we use Tensorflow Projector\footnote{https://projector.tensorflow.org/}.
We carry out the embedding vectors analysis using the vectors trained with data from 2010 to 2018 and testing with data from 2019 using LTF(2).

\begin{figure}	
	\caption{Embedding vector for the months variable using Tensorflow projector\label{fig:embedding month}}
	\includegraphics[viewport=0bp 460bp 500bp 780bp,scale=0.7]{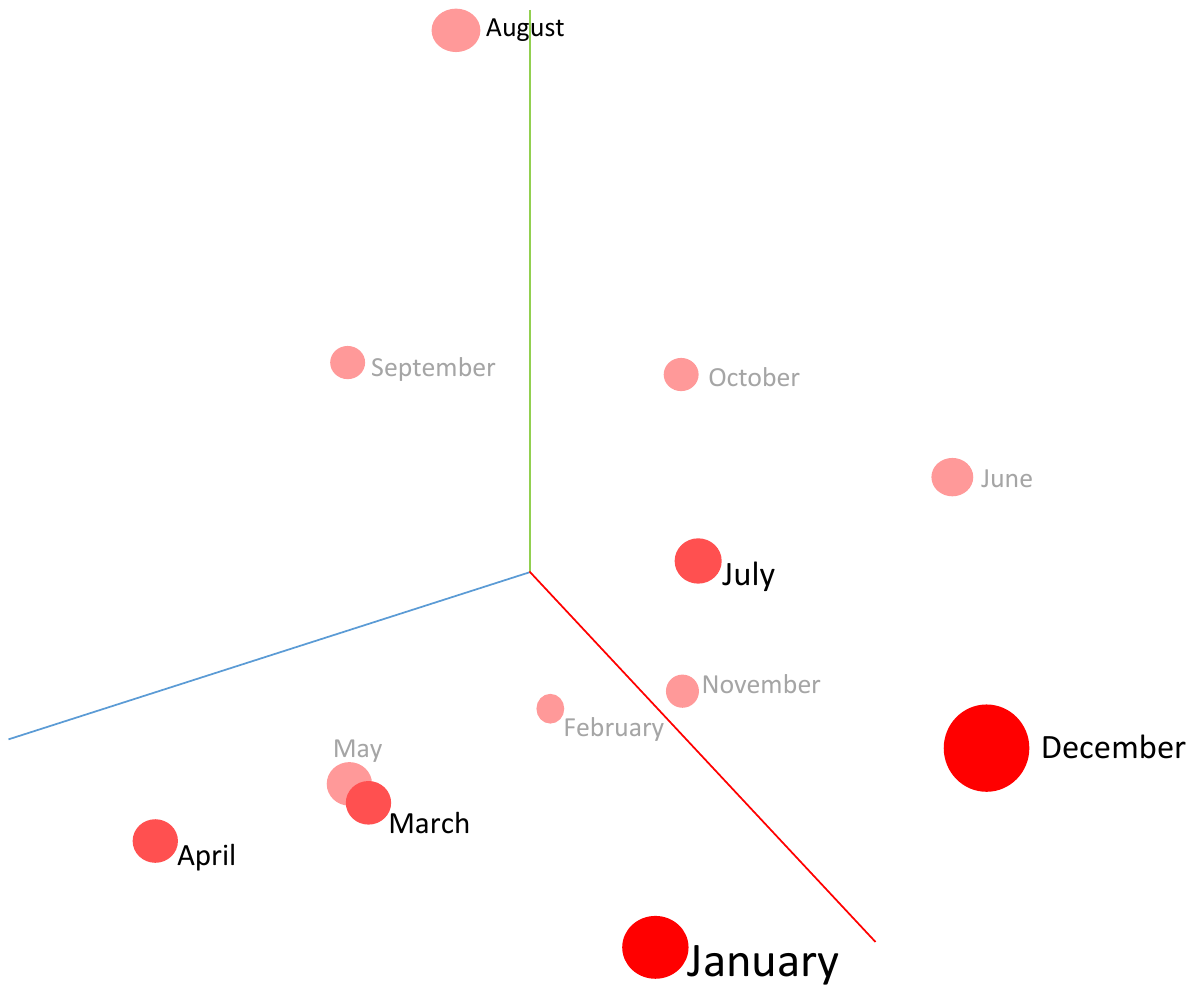}
\end{figure}

\begin{figure}
	\caption{Embedding vector for the weekday variable using Tensorflow projector\label{fig:embedding weekday}}
	\includegraphics[viewport=0bp 460bp 500bp 780bp,scale=0.7]{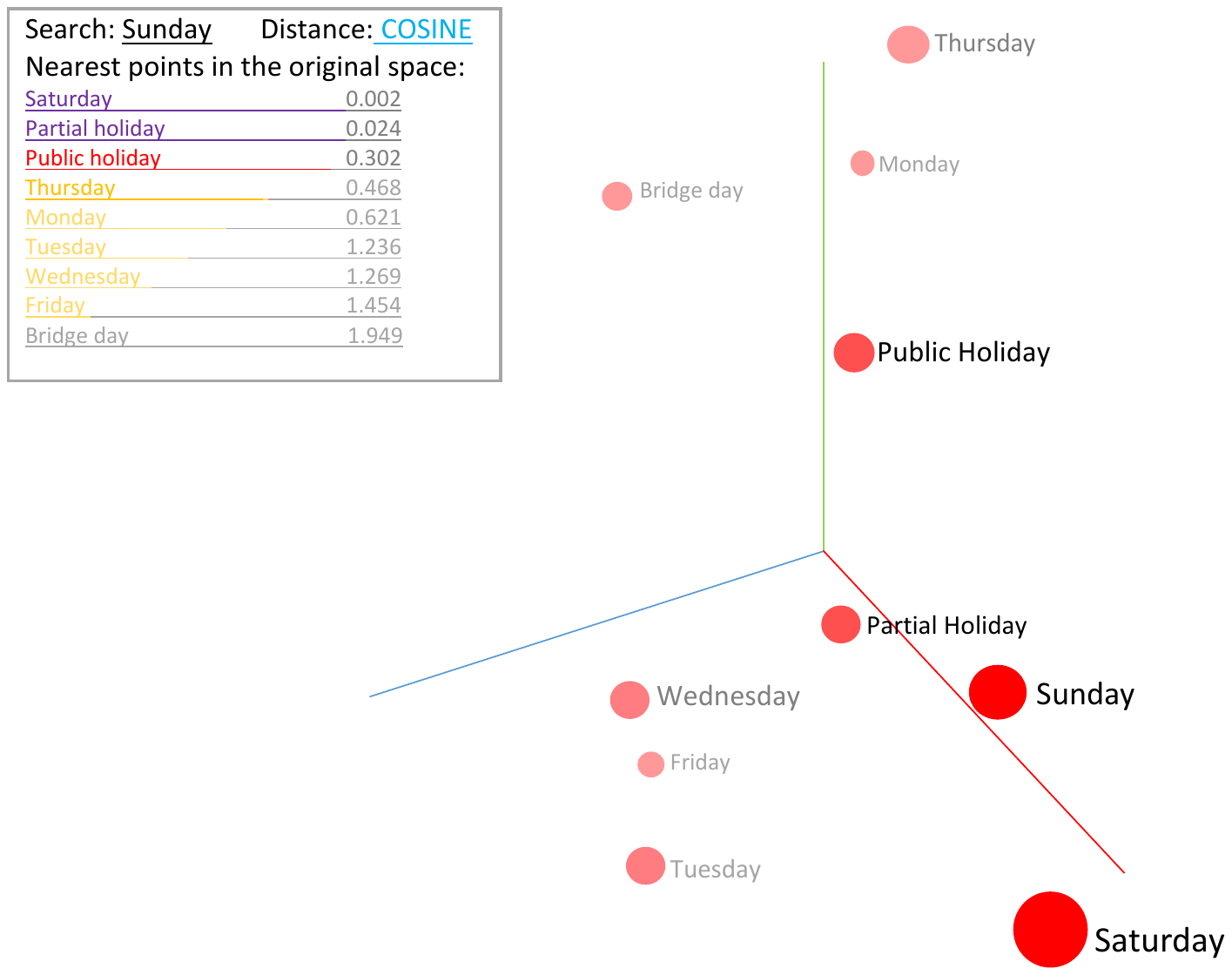}
\end{figure}

Figure \ref{fig:embedding month} shows a visualization of the variable month in the embedding space.
It has three dimensions, as explained in subsection \ref{subsec:proposed_emb}.
It can be seen that the embedding separates the winter months of January and December as well as August, which is the holiday season in Germany.
Also, the spring months March, April, and May are very close, similar to November and February, which are winter months without long periods of public holiday (in contrast to December and January).
Overall, the embedding shows a very reasonable picture from an energy economist's point-of-view.
Figure~\ref{fig:embedding weekday} shows the resulting visualization of the variable weekday in the embedding space.
We additionally show the use of the cosine-distance as a tool for the analysis of embeddings.
We calculate the distance of "Sunday" and find that it is very close to "Saturday", "partial holiday" and "public holiday", indicating that the network learned that the hourly profile is similar for those types of days.
This is again reasonable.\revTwoOut{ and would be confirmed by anyone working in electricity pricing.}

Although these findings seem to be well-known conclusions for electricity price experts, finding relationships between the days of the week, the months, and price variations is a very valuable tool in model analysis.
Neural networks typically are black-box models that hardly allow any insight into the model's logic.
This reduces the acceptance in practice, as there remains a significant risk, that the models "learned the wrong thing".
From our experience, one can reduce these risks by choosing simple network architectures and  extensive out-of-sample performance testing.
Both are fulfilled in this study, and additionally, we are able to provide graphical insight into the model's logic, which can be used to gain the trust of decision-makers.

\FloatBarrier

\section{Conclusion}\label{sec:concluion}
This paper provides a novel method to forecast electricity prices using machine learning,
\revTwo{which competes with existing approaches, but uses easier-to-understand model architectures, and provides insight into the model's logic.}
In an extensive study on the German electricity market, we \revTwo{compared our approach to benchmarks from the literature.}\revTwoOut{showed that the model produces nearly half the forecast error compared treo the existing default approach in short-term price forecasting on our dataset.}
We showed that the embedding approach can be used for the generation of long-term price profiles, as needed for the construction of hourly price forward curves.
Therefore utilities can base both applications on the same logic, which highly reduces the operational effort.
Additionally, we present tools to analyze the "black-box" neural network in order to reduce the model risk and to increase acceptance of our approach in practice.

The research questions have been answered.
Calendar information is \revTwoOut{best}\revTwo{well} included in neural networks using embeddings.\revTwoOut{ and this approach outperforms the \er{considered models in our experimental study}}. 
From the embeddings we can gain economic insights and analyze which calendar variables lead to similar behaviour of electricity prices.
\aw{
We believe that the presented approach of embedding calendar information can be used in many regression or classification tasks with a strong calendar dependency.
This includes potentially all electricity markets worldwide. Many other commodities show a seasonal behaviour in prices, however, such a dependency is usually not a feature in equity markets.
In a further research project we also successfully applied the proposed approach to the forecasting of air quality near roads (traffic also has a strong calendar dependency due to commuting patterns).
Therefore, future research could explore new applications of the calendar embedding, not only in price forecasting, but also in other domains.
Our technique, however, is not well suited for applications, which require not only solid forecasts but also deep model insights.
This is a natural limitation of all models using neural networks.
Further improvements of the forecasting quality of the model could be gained by adding additional fundamental factors (like system load or demand) or explicitly including a history of prices as features to account for its auto-regressive behaviour\footnote{We thank the anonymous referee for her comment.}.
}
\revTwo{There is also no evidence from our perspective that LSTM is a competitive approach in EPF, which is also stated in \cite{LAGO2021}: \textit{"Considering the most complete benchmark study in terms of forecasting models [57], it seems that a simple DNN with two layers is one of the best ML models. In particular, while more complex models, e.g. LSTMs, could potentially be more accurate, there is at the moment no sound evidence to validate this claim"}.
Further research may show if the proposed embedding layer and the benchmark models from \cite{LAGO2021} can be combined.}

\clearpage 
\section*{Acknowledgements}

This work has partially been supported by the German Federal Ministry for Economic Affairs and Energy in grant 01186724/1 (FlexEuro: Wirtschaftliche Optimierung flexibler stromintensiver Industrieprozesse) and the German Federal Ministry of Education and Research in grant 05M18AMC (ENets: Modellierung und Steuerung zukünftiger Energienetze). The research of Enislay Ramentol has been funded by the European Research Consortium for Informatics and Mathematics (ERCIM) Alain Bensoussan Fellowship Programme and the Fraunhofer Institute for Industrial Mathematics. The authors would like to thank Tania Jacob for the  valuable review of the manuscript.
We also thank the three anonymous referees for their valuable comments, which significantly improved the paper.

\begin{appendices}
\section{Detailed classification of public holidays}\label{app:holidays}
The embedding for calendar features builds upon the following classification of holidays:
\begin{itemize}
	\item \emph{public holidays}: Christmas, Day After Christmas, New Years Day, First of May (International Workers Day), Day of German Unity, Good Friday, Easter Sunday, Easter Monday, Ascension Day, Pentecost Monday
	\item \emph{partial holidays}: assumption of Mary, Reformation Day, All Hallows Day, Day of Prayer and Repentance, Pentecost Sunday, the Christmas week
	\item \emph{bridge days}: all days between public holidays, Fridays if Thursdays are public holidays and Mondays if Tuesdays are public holidays
\end{itemize}

\end{appendices}

\clearpage
\bibliographystyle{agsm}
\bibliography{references}

\end{document}